\documentclass[fleqn,usenatbib,useAMS]{mnras}

\usepackage{graphicx}	
\usepackage{amsmath}	
\usepackage{amssymb}	
\usepackage{multicol}        
\usepackage{bm}		
\usepackage{pdflscape}	
\usepackage{mathtools}
\usepackage{multirow}
\usepackage{xcolor}
\usepackage[utf8]{inputenc}
\usepackage[T1]{fontenc}
\usepackage{ae,aecompl}

\newcommand{\OmCIV}{\mbox{$\Omega_{\rm C \, \textsc{iv}}$}}
\newcommand{\OmCII}{\mbox{$\Omega_{\rm C \, \textsc{ii}}$}}
\newcommand{\HII}{\mbox{H \textsc{ii}}}

\newcommand{\CII}{\mbox{C \textsc{ii}}}
\newcommand{\CIII}{\mbox{C \textsc{iii}}}
\newcommand{\CIV}{\mbox{C \textsc{iv}}}

\newcommand{\SiII}{\mbox{Si \textsc{ii}}}
\newcommand{\SiIV}{\mbox{Si \textsc{iv}}}
\newcommand{\NV}{\mbox{N \textsc{v}}}
\newcommand{\HeII}{\mbox{He \textsc{ii}}}
\newcommand{\FeII}{\mbox{Fe \textsc{ii}}}
\newcommand{\MgII}{\mbox{Mg \textsc{ii}}}
\newcommand{\OI}{\mbox{O \textsc{i}}}
\newcommand{\Lya}{Ly$\alpha$}
\newcommand{\kms}{\,km\,s$^{-1}$}
\newcommand{\dndx}{d$n$/d$X$}

\defcitealias{Davies22Survey}{Paper~I}
\defcitealias{Harikane22a}{H22b}
\defcitealias{Madau14}{MD14}
\defcitealias{StorrieLombardi96}{SL96}

\newcommand{\appropto}{\mathrel{\vcenter{
  \offinterlineskip\halign{\hfil$##$\cr
    \propto\cr\noalign{\kern2pt}\sim\cr\noalign{\kern-2pt}}}}}

\newenvironment{nscenter}
 {\parskip=0pt\par\nopagebreak\centering}
 {\par\noindent\ignorespacesafterend}

\title[Redshift Evolution of C~IV Absorption using E-XQR-30]{Examining the Decline in the C~IV Content of the Universe over \mbox{4.3 $\lesssim z \lesssim$ 6.3} using the E-XQR-30 Sample}

\author[Rebecca L. Davies et al.]{\href{https://orcid.org/0000-0002-3324-4824}{Rebecca L. Davies}$^{1,2}$\thanks{Contact e-mail: \href{mailto:rdavies@swin.edu.au}{rdavies@swin.edu.au}}, %
\href{https://orcid.org/0000-0002-5360-8103}{E. Ryan-Weber}$^{1,2}$, %
\href{https://orcid.org/0000-0003-3693-3091}{V. D'Odorico}$^{3,4,5}$, %
\href{https://orcid.org/0000-0001-8582-7012}{S. E. I. Bosman}$^6$, %
\href{https://orcid.org/0000-0001-5492-4522}{R. A. Meyer}$^6$, %
\newauthor \href{https://orcid.org/0000-0003-2344-263X}{G. D. Becker}$^7$, %
\href{https://orcid.org/0000-0002-6830-9093}{G. Cupani}$^3$, %
\href{https://orcid.org/0000-0001-5211-1958}{L. C. Keating}$^{8}$, %
\href{https://orcid.org/0000-0002-4314-021X}{M. Bischetti}$^3$, %
\href{https://orcid.org/0000-0003-0821-3644}{F. B. Davies}$^6$, %
\href{https://orcid.org/0000-0003-2895-6218}{A.-C.~Eilers}$^9$\textdagger, 
\newauthor \href{0000-0002-6822-2254}{E. P. Farina}$^{10}$, %
\href{https://orcid.org/0000-0001-8443-2393}{M. G. Haehnelt}$^{11,12}$, %
\href{https://orcid.org/0000-0002-7129-5761}{A. Pallottini}$^{5}$, %
\href{0000-0003-3307-7525}{Y. Zhu}$^7$ \\
$^1$Centre for Astrophysics and Supercomputing, Swinburne University of Technology, Hawthorn, Victoria 3122, Australia \\
$^2$ARC Centre of Excellence for All Sky Astrophysics in 3 Dimensions (ASTRO 3D), Australia \\
$^3$INAF-Osservatorio Astronomico di Trieste, Via Tiepolo 11, I-34143 Trieste, Italy \\
$^4$IFPU-Institute for Fundamental Physics of the Universe, via Beirut 2, I-34151 Trieste, Italy \\
$^5$Scuola Normale Superiore, Piazza dei Cavalieri 7, I-56126 Pisa, Italy \\
$^6$Max-Planck-Institut f\"ur Astronomie, K\"onigstuhl 17, D-69117 Heidelberg, Germany \\
$^7$Department of Physics \& Astronomy, University of California, Riverside, CA 92521, USA \\
$^8$Leibniz Institute for Astrophysics Potsdam (AIP), An der Sternwarte 16, D-14482 Potsdam, Germany \\
$^9$MIT Kavli Institute for Astrophysics and Space Research, 77 Massachusetts Ave., Cambridge, MA 02139, USA \\
$^{10}$Gemini Observatory, NSF’s NOIRLab, 670 N A’ohoku Place, Hilo, Hawai'i 96720, USA \\
$^{11}$Institute of Astronomy, University of Cambridge, Madingley Road, Cambridge CB3 0HA, UK \\
$^{12}$Kavli Institute for Cosmology, University of Cambridge, Madingley Road, Cambridge CB3 0HA, UK \\
\textdagger NASA Hubble Fellow\\
}

\pubyear{2022}

\begin{document}
\label{firstpage}
\pagerange{\pageref{firstpage}--\pageref{lastpage}}
\maketitle

\begin{abstract}
Intervening \CIV\ absorbers are key tracers of metal-enriched gas in galaxy halos over cosmic time. Previous studies suggest that the \CIV\ cosmic mass density (\OmCIV) decreases slowly over \mbox{1.5~$\lesssim\,z\lesssim$~5} before declining rapidly at \mbox{$z\gtrsim$~5}, but the cause of this downturn is poorly understood. We characterize the \OmCIV\ evolution over \mbox{4.3~$\lesssim z\lesssim$~6.3} using 260 absorbers found in 42 XSHOOTER spectra of $z\sim$~6 quasars, of which 30 come from the ESO Large Program XQR-30. The large sample enables us to robustly constrain the rate and timing of the downturn. We find that \OmCIV\ decreases by a factor of 4.8~$\pm$~2.0 over the \mbox{$\sim$~300 Myr} interval between $z\sim$~4.7 and $z\sim$~5.8. The slope of the column density ($\log N$) distribution function does not change, suggesting that \CIV\ absorption is suppressed approximately uniformly across \mbox{13.2~$\leq\log N$/cm$^{-2}$~$<$~15.0}. Assuming that the carbon content of galaxy halos evolves as the integral of the cosmic star formation rate density (with some delay due to stellar lifetimes and outflow travel times), we show that chemical evolution alone could plausibly explain the fast decline in \OmCIV\ over \mbox{4.3~$\lesssim z\lesssim$~6.3}. However, the \CIV/\CII\ ratio decreases at the highest redshifts, so the accelerated decline in \OmCIV\ at $z\gtrsim$~5 may be more naturally explained by rapid changes in the gas ionization state driven by evolution of the UV background towards the end of hydrogen reionization.
\end{abstract}


\section{Introduction}
The formation of the first galaxies marked an important turning point in cosmic history. Massive stars released high energy photons which commenced the reionization of the Universe, and stellar nucleosynthesis led to the production of the first heavy elements which were then released into the surrounding gas via supernova explosions. However, relatively little is known about the timing of the formation of the first galaxies and how they shaped the properties of their surrounding environments \citep[see][for reviews]{Bromm11, Dayal18, Robertson21}. The shortcomings in our understanding have been emphasized by recent observational claims of unexpectedly massive galaxies at $z\gtrsim$~10 revealed by JWST \citep[e.g.][]{Atek22, Furtak22, Harikane22b, Labbe22, Naidu22, Yan22, Adams23}.

Our understanding of the properties of the Universe near the end of the Epoch of Reionization has been propelled by the detection of growing numbers of $z\sim$~6 quasars. Constraints on the timing of reionization from measurements of \Lya\ forest dark gaps \citep[e.g.][]{Zhu21, Zhu22} and transmission statistics \citep[e.g.][]{Fan06, Becker15, Bosman18, Eilers18, Yang20, Bosman22} and \Lya\ damping wing absorption \citep[e.g.][]{DaviesF18, Greig22} are consistent with leading independent probes of reionization (e.g. the luminosity function of \Lya\ emitters; \citealt{Konno18}, the fraction of Lyman Break galaxies showing \Lya\ emission; \citealt{Mason18}, and the optical depth to reionization; \citealt{Planck18}) and suggest that signatures of reionization persist below $z\sim$~6. Furthermore, intervening metal absorption lines in $z\sim$~6 quasar spectra provide key insights into the chemical content and ionization state of gas around early galaxies. The optical depths of absorption lines remain constant as the quasar light travels through space, making it possible to investigate the properties of faint low-mass galaxies that fall below the detection limits of current emission-line surveys. The comoving mass densities of metal ions trace the overall chemical content of the Universe over cosmic time, whilst ionic ratios provide constraints on the ionization state of the absorbing gas (see e.g. \citealt{Becker15Review} for a review).

The \mbox{\CIV~$\lambda \lambda$~1548, 1550\AA\ doublet} is commonly used as a tracer of enriched gas across cosmic time because it is observable across a wide redshift range and is easily identifiable due to its doublet nature. \CIV\ is produced by photons with an energy of at least 47.9 eV and primarily traces metals in the circumgalactic and intergalactic media (CGM and IGM), observed as absorption systems \citep[e.g.][]{Schaye03, Pettini03, Peroux04, Simcoe04, Songaila05, Songaila06, Simcoe06, Schaye07, Danforth08, Becker09, RyanWeber09, DOdorico10, Simcoe11, Tilton12, Cooksey13, DOdorico13, Shull14, Boksenberg15, Burchett15, Danforth16, Diaz16, DOdorico16, Kim16, Bosman17, Codoreanu18, Cooper19, Hasan20, Manuwal21, Hasan22} as well as emission nebulae around quasars \citep[e.g.][]{Guo20, Travascio20}. 

Leading theories of CGM and IGM enrichment suggest that the majority of the carbon traced by \CIV\ absorbers was not formed in-situ, but was produced in stars and subsequently ejected from galaxies by means of outflows \citep[e.g.][]{Aguirre01, Theuns02, Oppenheimer06, Kobayashi07, Oppenheimer09, Cen11, Finlator13, Finlator20, Yamaguchi22}. At $z\gtrsim$~5, star-formation driven outflows are expected to be the dominant source of metal-enrichment \citep[e.g.][]{Tescari11, Pallottini14, Suresh15, Sorini20}. Measurements of the statistics of \CIV\ absorbers at $z\gtrsim$~5 therefore provide valuable constraints on sub-grid feedback models, which are typically tuned to reproduce the star formation rate and stellar mass content of galaxies but not the CGM properties \citep[e.g.][]{Oppenheimer06, Suresh15, Keating16, Rahmati16, Garcia17b, Finlator20}. 

Previous studies of \CIV\ absorption across cosmic time have found that the comoving mass density of \CIV\ (\OmCIV) declines smoothly at \mbox{1.5 $\lesssim z \lesssim$ 5} before dropping rapidly at $z\gtrsim$~5 \citep[e.g.][]{Songaila97, Songaila01, Songaila05, Becker09, RyanWeber09, DOdorico10, Simcoe11, DOdorico13, Boksenberg15, Diaz16, Bosman17, Codoreanu18, Meyer19, DOdorico22}. This could be an indication that the carbon content of galaxy halos grows quickly over \mbox{6 $\gtrsim z \gtrsim$ 5} due to early enrichment by outflows. However, \OmCIV\ also depends on the fraction of carbon existing as \CIV\ which is determined by the ionization state of the CGM/IGM. There is growing observational evidence for a rapid transition in the typical ionization environments of metal absorbers at $z\sim$~5.7. The observed decline in \CIV\ absorption is mirrored in \mbox{\SiIV~$\lambda \lambda$1393, 1402\AA} \citep[e.g.][]{DOdorico22} which is produced by photons of similar energy to \CIV\ ($\geq$~33.5~eV). In contrast, the incidence of weak low-ionization absorbers (e.g. \mbox{\MgII~$\lambda \lambda$2796, 2803\AA} and \mbox{\OI~$\lambda$1302\AA}) remains constant or increases at $z\gtrsim$~5.7 \citep[e.g.][]{Bosman17, Chen17, Codoreanu17, Becker19}. The ratio of \CIV\ to \mbox{\CII~$\lambda$1334\AA}, which probes the distribution of carbon between ionization states, drops at $z \gtrsim$~5.7 \citep[e.g.][]{Cooper19}. \CIV\ absorption is often weak or undetected in low-ionization metal absorbers at $z\sim$~6 \citep[e.g.][]{DOdorico18, Simcoe20}, unlike similar systems at \mbox{$z\sim$~2~--~3} which show ubiquitous \CIV\ absorption \citep[e.g.][]{Songaila06, Prochaska13, Rubin15, Cooper19}. These findings suggest that there is a transition in the average ionization environments of metal absorbers, from more neutral and low-ionization environments at $z\sim$~6 to more highly ionized environments at $z\lesssim$~5. This transition may be driven by an increase in the hardness and/or amplitude of the UV background following the end of reionization \citep[e.g.][]{Oppenheimer09, Finlator15, Becker19}.

Large quasar samples are required to robustly examine the relative contributions of chemical enrichment and changes in ionization state to the rapid evolution in \OmCIV\ over \mbox{5 $\lesssim z \lesssim$~6}. The Ultimate XSHOOTER legacy survey of quasars at \mbox{$z\sim$~5.8~--~6.6} (XQR-30; D'Odorico et al. in prep) has obtained 30 high signal-to-noise (S/N) quasar spectra at $\sim$~30~\kms\ resolution, nearly quadrupling the previous sample of 12 high quality $z\sim$~6 quasar spectra in the XSHOOTER archive. The combined sample of 42 spectra is referred to as the enlarged \mbox{XQR-30} or \mbox{E-XQR-30} sample. The first publications based on this sample have revealed that relativistic quasar-driven outflows are much more prevalent at $z\sim$~6 than at $z\sim$~2~--~4 \citep{Bischetti22}, that fluctuations related to re-ionization persist in the IGM until at least $z$~=~5.3 \citep{Zhu21, Bosman22}, that bright $z\sim$~6 quasars typically live in overdense environments \citep{Chen22}, and that quasar broad line regions at $z\sim$~6 are already enriched to $\sim$~2~--~4 times the solar abundance \citep{Lai22}.

\citet[][hereafter \citetalias{Davies22Survey}]{Davies22Survey} performed a systematic search for metal absorption lines in the E-XQR-30 spectra and published a catalog of identified absorption systems. We use a sample of 260 intervening \CIV\ absorbers from this catalog to make robust measurements of the number density, cosmic mass density, and column density distribution function of \CIV\ absorbers over \mbox{4.3 $\lesssim z \lesssim$ 6.3}. Using these measurements, we examine when and how rapidly the drop in \OmCIV\ occurs, and explore whether this decline is driven primarily by chemical evolution and/or a change in the fraction of carbon found as \CIV. 

The paper is structured as follows. We describe the metal absorber catalog and the properties of the \CIV\ absorber sample in Section \ref{sec:sample}. The \CIV\ line statistics are presented and compared to predictions from the \textit{Technicolor Dawn} simulation in Section \ref{sec:results}. We examine the contributions of chemical evolution and changes in ionization state to the \OmCIV\ evolution in Section \ref{sec:discussion}, and summarize our results in Section \ref{sec:conclusions}.

Throughout this work we adopt the \citet{Planck18} $\Lambda$CDM cosmology with \mbox{H$_{0}$ = 67.7 \kms\ Mpc$^{-1}$} and \mbox{$\Omega_m$ = 0.31}. 

\section{Sample and Data Processing}\label{sec:sample}
\subsection{Quasar Spectra}
The sample of \CIV\ absorbers used in this paper is drawn from the publicly available XQR-30 metal absorber catalog\footnote{Available on GitHub: \href{https://github.com/XQR-30/Metal-catalogue}{\url{https://github.com/XQR-30/Metal-catalogue}}}, which was assembled through a systematic search for absorption lines in the spectra of 42 $z\sim$~6 quasars \citepalias{Davies22Survey}. 30 of the quasars were observed as part of XQR-30, an ESO Large Program (PI: V. D'Odorico) which obtained deep (continuum S/N ratio $\gtrsim$~10 per 10~\kms\ spectral pixel at \mbox{1285 \AA} rest-frame), medium resolution (\mbox{R $\simeq$~10,000}, \mbox{FWHM $\simeq$ 30~\kms}) spectra of luminous \mbox{(J$_{\rm AB} \le 19.8$)} quasars at \mbox{5.8 $\lesssim z \lesssim$~6.6} using XSHOOTER \citep{Vernet11}. The observations were performed using slit widths of 0.9'' and 0.6'' for the VIS and NIR spectroscopic arms, respectively. The remaining 12 spectra were sourced from archival XSHOOTER observations of quasars in the same redshift and $J$-band magnitude range that were observed at similar spectral resolution and meet the S/N requirements for XQR-30. The full dataset is described in detail in D'Odorico et al. (in prep). All quasars were selected based on their redshift and $J$-band magnitude, with no prior knowledge of intervening absorber properties.

The XQR-30 and archival observations were reduced using the same pipeline for consistency. The data reduction process is outlined in \citet{Becker19}. Briefly, a composite dark frame was subtracted from each exposure, sky subtraction was performed on the un-rectified frame \citep{Kelson03}, and an initial 1D spectrum was extracted using the optimal weighting method of \citet{Horne86}. The Cerro Paranal Advanced Sky Model \citep{Noll12, Jones13} was used to compute telluric corrections which were then applied back to the 2D spectra. To maximize the bad pixel rejection efficiency, all exposures were processed simultaneously when extracting the final 1D spectrum for each quasar and XSHOOTER spectroscopic arm (VIS and NIR). The final extracted spectra have a velocity sampling of 10~\kms.

The widths of the absorption line profiles revealed that when the average seeing is significantly smaller than the slit width, the true spectral resolution can be significantly higher than the nominal (slit-width-dependent) value. The true spectral resolution of each extracted spectrum was estimated as described in D'Odorico et al. (in prep). The average widths of spectral order spatial profiles were measured for individual high S/N 2D frames (a single value per exposure) and used to derive empirical relationships between the average spectral resolution and the seeing for each XSHOOTER arm and slit width. The empirical relationships were then used to convert the recorded seeing values into estimates of the Gaussian width of the line spread function (LSF) for each exposure. The final spectral resolution for each spectrum was obtained by computing the (S/N)$^2$-weighted average of the individual LSFs. The adopted spectral resolutions are listed in \citetalias{Davies22Survey} and range from \mbox{$R$ = 9500~--~13700} (median 11400, corresponding to \mbox{FWHM = 26~\kms}) for the VIS arm and \mbox{$R$ = 7600~--~11000} (median 9750, corresponding to \mbox{FWHM = 31~\kms}) for the NIR arm.

Finally, the VIS and NIR spectra were combined to produce a single spectrum for each quasar. The NIR spectrum was scaled to match the flux level of the VIS spectrum over \mbox{990~--~1015nm} and both spectra were cut at 1015nm before being stitched together, manually correcting for small \mbox{($\Delta v \simeq$~5~--~20~\kms)} mismatches between the relative wavelength calibrations of the arms as necessary. We note that the absolute flux scaling of the combined spectrum does not impact the derived absorption line properties which are measured from continuum-normalized spectra. The continuum fitting was performed using univariate spline interpolation within the \textsc{Python} library \textsc{astrocook}\footnote{\href{https://github.com/DAS-OATs/astrocook}{\url{https://github.com/DAS-OATs/astrocook}}} \citep{Cupani20}, as described in \citetalias{Davies22Survey}. The continuum fits are publicly available at the same location as the metal absorber catalog (see footnote 1).

\subsection{Metal Absorber Catalog}\label{subsec:catalog}
The 42 quasar spectra were systematically searched for metal absorption lines as described in \citetalias{Davies22Survey}. Briefly, an initial candidate list was generated by peforming an automated search for systems showing absorption in \MgII, \FeII\ ($\lambda$2344, 2382, 2586, and/or 2600\AA), \CII\ (in conjunction with other low-ionization lines), \CIV, \SiIV, and/or \NV\ ($\lambda \lambda$1238, 1242\AA) using \textsc{astrocook}. Only the wavelength regions redward of the quasar \Lya\ emission lines were searched because the saturation of the $z\sim$~6 \Lya\ forest makes it impossible to identify individual absorbers at shorter wavelengths. When performing this search, wavelength regions showing particularly strong skyline or telluric residuals in individual quasar spectra (most commonly the regions between the $J$, $H$, and $K$ observing bands at 1.35~--~1.45$\mu$m and 1.8~--~1.95$\mu$m) were masked to prevent the algorithm from returning a large number of spurious candidates. We note that approximately half of the spectra show broad absorption line (BAL) features associated with quasar-driven outflows (\citealt{Bischetti22} and submitted). The BAL regions were not masked, but any absorbers found in these regions were flagged in the final catalog.

\begin{figure*}
 \includegraphics[scale=0.55, clip = True, trim = 0 0 5 0]{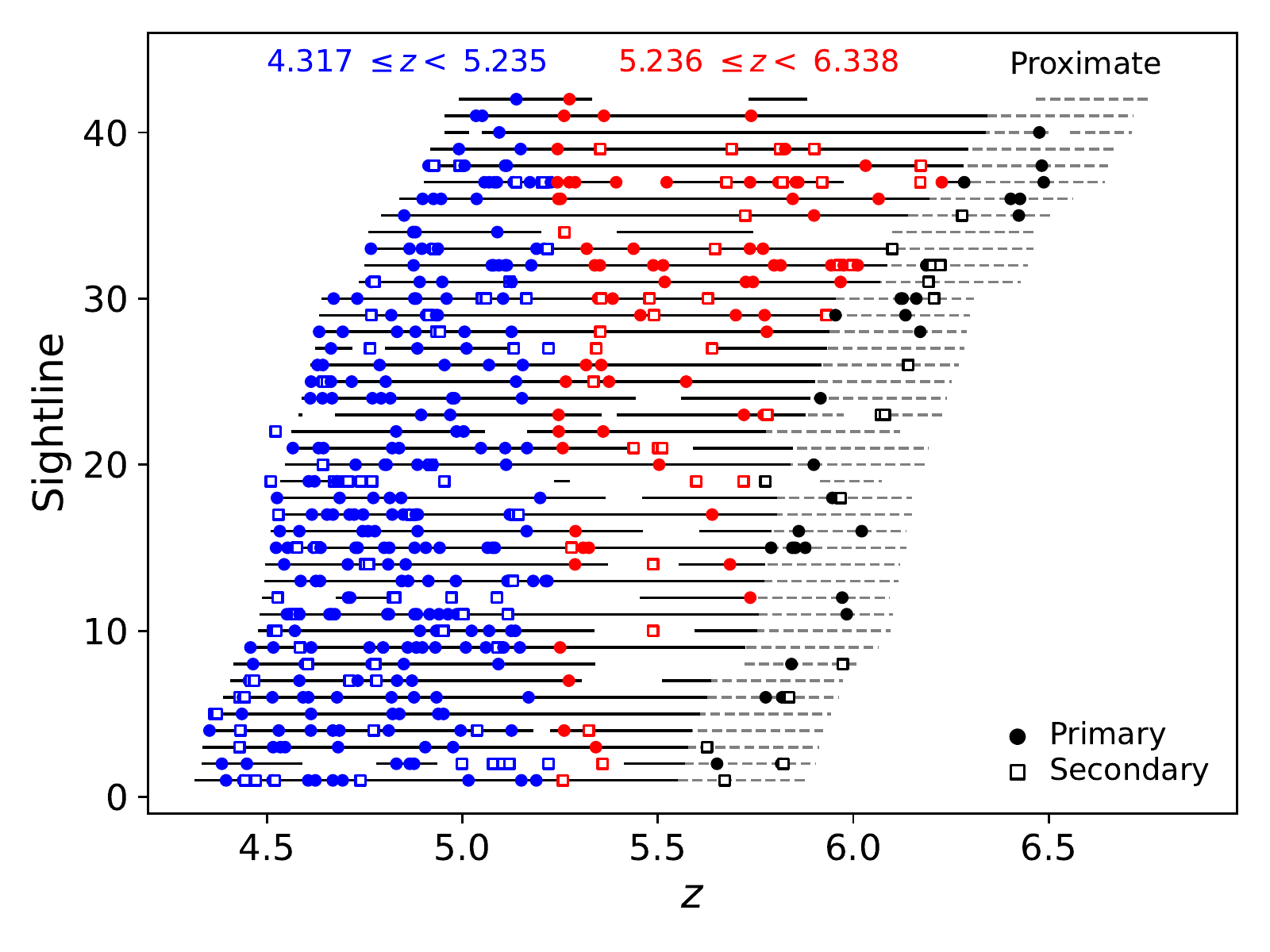}  \includegraphics[scale=0.55, clip = True, trim = 5 0 0 0]{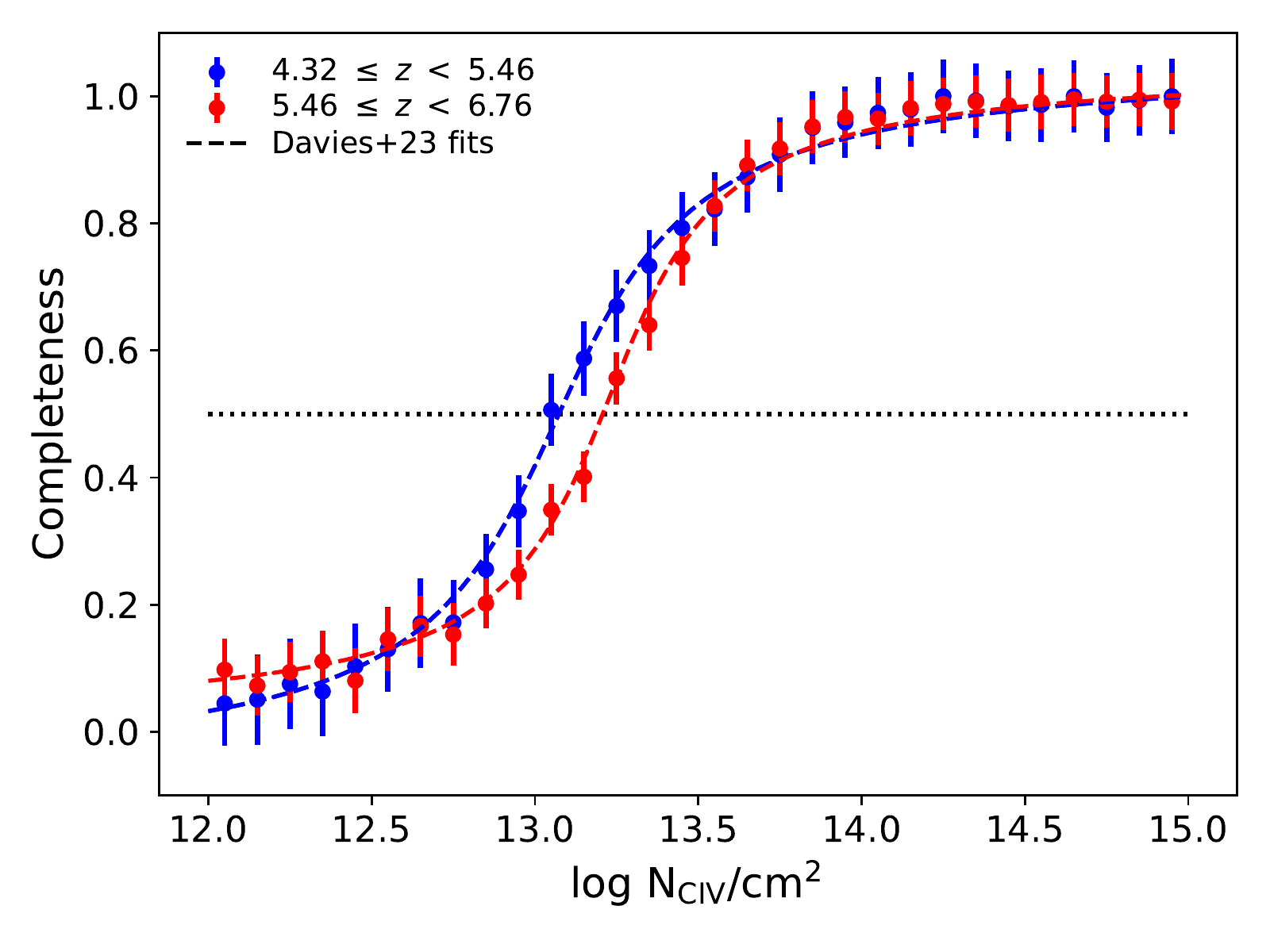}
\caption{Left: Illustration of the \CIV\ absorber sample. Horizontal lines show the redshift intervals over which the search for \CIV\ absorbers was conducted for each of the 42 quasar sightlines. Grey dashed regions highlight proximity zones within 10,000~\kms\ of the quasar redshift. Longer gaps trace redshift intervals where the \CIV\ lines fall within BAL features or regions that were masked due to strong skyline or telluric contamination. The markers show all \CIV\ absorbers in the catalog, where solid circles indicate primary absorbers (those that were automatically identified, pass the visual inspection check, and do not fall in masked wavelength regions or BAL regions) and open squares indicate secondary absorbers (all others). Proximate absorbers are shown in black. The intervening absorbers are split into two redshift bins (indicated by the marker color), divided at the path-length-weighted mean redshift of our survey. Right: Completeness as a function of column density for \CIV\ absorbers in the two redshift intervals considered in \citetalias{Davies22Survey}. The dashed lines show the best-fit arctan functions published in that work.} \label{fig:completeness} 
\end{figure*}

The list of candidate systems was automatically filtered to remove spurious systems originating from strong skyline residuals or chance alignment of unassociated transitions in velocity space. For each of the remaining candidates, an automated search was performed to identify potential absorption in additional ions and/or transitions at the same redshift. Regions showing strong skyline/telluric residuals were not masked during this search because the redshift prior set by the initial system significantly increases the probability that any aligned absorption is real. The candidate systems were visually inspected by five experts to remove remaining spurious systems. After this initial line identification phase was completed, the spectra were searched for unidentified absorption features. In cases where unidentified absorption lines could be visually identified, the relevant transitions/systems were added to the catalog with flags indicating that they were not recovered by the automatic line finder. 

The final metal absorber catalog contains a total of 727 \CIV\ absorption components spanning the redshift range \mbox{4.353 $\leq z \leq$ 6.487}. The physical properties ($z$, Dopper $b$ parameter, and column density, $\log N\equiv \log N$/cm$^{-2}$) of all absorption components were measured by fitting Voigt profiles using the \textsc{Astrocook} GUI. 

\subsection{Sample Selection}\label{subsec:sample}
In order to accurately measure the evolution in \CIV\ absorber properties over cosmic time we require robust estimates of the sample completeness (the fraction of true systems that are recovered) and the false-positive rate (the fraction of recovered systems that are spurious) as a function of $\log N$. The calculations are described in detail in \citetalias{Davies22Survey}. A set of 840 mock spectra (20 per quasar) was generated with known absorber properties (spanning a wide range in redshift, $\log N$ and $b$ parameter). The mock spectra were processed using the same steps applied to the observed spectra to generate mock absorber catalogs which were then compared to the lists of inserted systems. The only exception is that no attempt was made to visually determine the origin of unidentified absorption features in the mock spectra. Therefore, the completeness and false-positive rate are only well defined for systems that were found in the automated line search and pass the visual inspection check. The completeness cannot be robustly quantified in BAL regions because the broad absorption makes it more difficult to recover underlying narrow absorption. The completeness in the masked wavelength regions is poorly defined because these regions were excluded from the initial line search.

For these reasons, the metal absorber catalog is divided into a primary sample which contains absorbers that were automatically identified, pass the visual inspection check, and do not fall in masked wavelength regions or BAL regions, and a secondary sample containing absorbers that fail one or more of these criteria. In this study we only use the primary absorber sample to ensure that the completeness corrections are robust\footnote{We examined the impact of this choice on the derived \CIV\ absorber statistics by measuring the number densities and cosmic mass densities of strong \CIV\ absorbers (\mbox{13.8 $\leq \log N <$ 15.0}, for which the completeness corrections are neglible) i) using only the primary sample and ii) including manually identified systems that do not lie in masked/BAL regions. The two sets of measurements differ by $<$~15\% and are consistent within the 1$\sigma$ errors.}.

In the left hand panel of Figure \ref{fig:completeness}, the horizontal lines show the redshift ranges over which the search for \CIV\ absorption was conducted for each of the 42 quasar sightlines. The grey dashed regions highlight the proximity zones (within 10,000~\kms\ of the quasar redshifts), and longer gaps trace redshift intervals where \CIV\ falls within BAL features or regions that were masked due to strong skyline or telluric contamination. The markers show all of the \CIV\ systems in the catalog, where solid circles and open squares represent absorbers that are part of the primary and secondary sample,  respectively. Of the 727 \CIV\ components in the complete catalog, 559 are part of the primary sample. 

We additionally restrict our sample to intervening absorbers (which are conservatively defined to have a velocity offset of \mbox{$>$~10,000~\kms} from the quasar redshift) because absorbers in close proximity to quasars may not reflect the average ionization state and abundance patterns of the underlying absorber population \citep[e.g.][]{Berg16,Perrotta16}. Proximate absorbers are shown in black, while the intervening absorbers are shown in blue and red (where the color indicates whether the absorber lies above or below the path-length-weighted mean redshift of the survey; see Section \ref{subsec:absorption_path}). Of the 559 primary \CIV\ absorbers, 507 are intervening. 

\citetalias{Davies22Survey} showed that the false-positive rate in the catalog is negligible ($<$~5\%) at all column densities, and the completeness as a function of $\log N$ is well described by an arctan function. The best-fit arctan parameters were provided in two redshift bins (\mbox{4.32 $\leq z <$ 5.46} and \mbox{5.46 $\leq z <$ 6.76})\footnote{$z$~=~5.46 is the path-length-weighted mean redshift of the survey when quasar proximity zones are included.} as shown in the right-hand panel of Figure \ref{fig:completeness}. The \CIV\ sample is 50\% complete for \mbox{$\log N \geq$ 13.08} (\mbox{$\log N \geq$ 13.22}) at \mbox{4.32 $\leq z <$ 5.45} (\mbox{5.45 $\leq z <$ 6.76}, see Table 2 of \citetalias{Davies22Survey}). The survey is slightly less sensitive to \CIV\ absorption in the higher redshift range because the typical noise level in the spectra increases towards longer wavelengths. In our analysis we only include components with column densities above the 50\% completeness limit of the higher redshift bin (\mbox{$\log N \geq$~13.2}) because the number of detected systems drops at lower column densities, increasing the sampling error and reducing the reliability of the measurements. This cut reduces the sample from 507 components to 274. 

Finally, we exclude components with $\log N >$~15 to be consistent with previous works. This criterion results in the removal of a single component at \mbox{$z$~=~5.109} in VDESJ0224-4711. Our final sample consists of 273 \CIV\ absorption components spanning the redshift range \mbox{4.353 $\leq z \leq$ 6.067}. Table \ref{table:sample_selection} summarizes the number of absorption components from the original absorber catalog that meet successive selection criteria. All components in our final sample are detected at $\geq$~3$\sigma$ significance, and none are strongly saturated (i.e. the central optical depth of \CIV~$\lambda$1550 is \mbox{$\tau <$~2}). 

\begin{table}
\begin{nscenter}
\begin{tabular}{lc}
\hline
Criterion & Number of Components \\
\hline
\CIV\ & 727 \\
+ Primary Sample & 559 \\
+ Intervening & 507 \\
+ 13.2 $\leq \log N <$ 15.0 & 273 \\
\hline
\end{tabular}
\caption{The number of absorption components from the original absorber catalog that meet successive selection criteria, as described in Section \ref{subsec:sample}.}\label{table:sample_selection}
\end{nscenter}
\end{table}

\subsection{Grouping Components}
The \CIV\ absorber statistics can be calculated using either individual Voigt profile components of the recovered \CIV\ absorbers, or systems that group all components lying within a chosen velocity interval $\Delta v$. The choice of whether to use components or systems directly impacts the measured number density and the slope of the column density distribution function \cite[e.g.][]{Boksenberg15}. The latter translates into a minor impact on \OmCIV\ because it is measured over a specific $\log N$ range (see Section \ref{subsec:om_civ}).

In this work, we choose to examine the statistics of individual components. The number density and column density distribution function of individual components are strongly resolution-dependent because components separated by less than the spectral resolution may be blended. In Section \ref{sec:results}, we directly compare our \CIV\ absorber statistics with those of lower redshift absorbers identified in much higher resolution spectra \mbox{($R \simeq$~50,000)}. To enable consistent comparison of all datasets, we merge components with $\Delta v <$~50~\kms\ (this threshold is commonly used because it is the coarsest spectral resolution for typical absorption line measurements). Merged components are combined into a single `system' defined by the total column density and column-density-weighted mean redshift of the constituent components \citep{Songaila01, DOdorico10}. We emphasize that the merging is not intended to group all physically associated absorption components, but rather provides a means to consistently compare datasets that have significantly different spectral resolutions. The 273 \CIV\ components in our final sample comprise a total of 260 systems.

\subsection{Comparison with Previous Catalogs}
Catalogs of \CIV\ absorption have been previously published for 12/42 of the E-XQR-30 quasars, based on either archival spectra or lower quality spectra of the XQR-30 quasars \citep{Codoreanu18, Meyer19, DOdorico13, DOdorico22}. A detailed comparison with these catalogs is given in Section 5.4 of \citetalias{Davies22Survey}. In brief, we recover 107/122 (88\%) of the previously reported \CIV\ systems. 4/15 (27\%) of the missed systems were found to be better explained by transitions of other ions at different redshifts. Our comprehensive, simultaneous search for low and high-ionization systems allows us to more robustly identify the origin(s) of individual absorption lines compared to searches focused on individual ions. One system (7\%) was rejected by the checkers due to strong skyline contamination. The remaining 10/15 (60\%) missed systems have column densities near or below the completeness limit of our data and were missed due to low S/N or blending with much stronger absorption features. \citet{DOdorico13} performed line identification visually and \citet{DOdorico22} filtered candidates manually rather than using an automated algorithm, and this likely explains why they were able to detect some weak systems that were missed in our catalog. However, the improved quality of the \mbox{E-XQR-30} spectra enabled the detection of approximately 50 new \CIV\ systems along the lines of sight to the same quasars, increasing the number of known systems in these spectra by $\sim$~30\%.

\subsection{Absorption Path}\label{subsec:absorption_path}
The first step in calculating the incidence of \CIV\ absorbers over a given redshift interval is to compute the total absorption search path covered by the survey. The absorption distance to an object at a redshift of $z$ is given by:
\begin{equation}
 X(z) = \frac{2}{3 \Omega_m} \left(\Omega_m \left(1 + z\right)^3 + \Omega_\Lambda \right)^{1/2}
\end{equation}
\citep{Bahcall69}. 
The absorption search path covered by a single sightline for which \CIV\ is accessible over the range \mbox{$z_1 \leq z < z_2$} is given by $\Delta X = X(z_1) - X(z_2)$. If the redshift search interval is broken up by masked or BAL regions, then $\Delta X$ should be computed for each sub-interval individually. The total absorption path covered by our sample of 42 quasar spectra over a redshift range \mbox{$z_1 \leq z < z_2$} can be calculated by taking the redshift search intervals shown by the black horizontal lines in the left-hand panel of Figure \ref{fig:completeness} and summing $\Delta X$ for all sub-intervals that lie within that redshift range. The redshift search intervals can be reproduced using the data in Tables B1, B2 and B4 of \citetalias{Davies22Survey} which list the adopted quasar redshifts, the masked wavelength regions, and the BAL regions (originally published by \citealt{Bischetti22} and Bischetti et al. submitted), respectively. The absorption path length values quoted in this paper can be recovered using the \textsc{python} code\footnote{The routine can be used to calculate the absorption path for the public E-XQR-30 metal absorber catalog over a given redshift interval for a range of ions/transitions and any value of $\Omega_m$.} and data provided with the metal absorber catalog (see footnote 1).

\begin{figure}
\includegraphics[scale=0.55, clip = True, trim = 0 0 5 0]{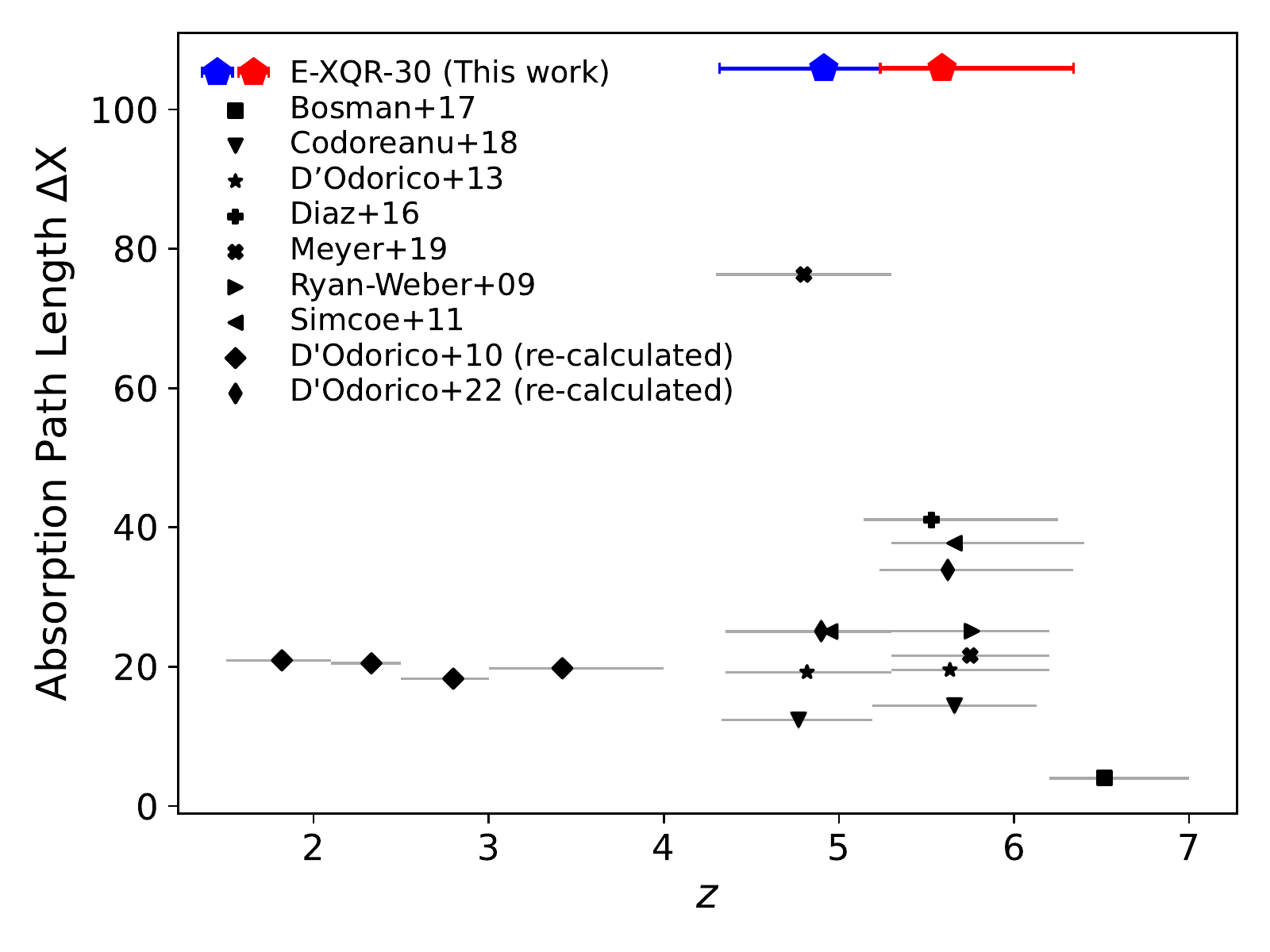} 
\caption{The absorption path length for intervening \CIV\ absorbers covered by the E-XQR-30 survey over the two redshift bins shown in the left-hand panel of Figure \ref{fig:completeness} (blue and red pentagons). Black markers show the path lengths covered by other surveys of \CIV\ absorption in the literature. The literature samples differ significantly from one another in the column density thresholds above which they are complete.} \label{fig:path_length_comparison} 
\end{figure}

Figure \ref{fig:path_length_comparison} compares the absorption path length for intervening \CIV\ absorbers covered by the E-XQR-30 survey to those of previous surveys of \CIV\ absorption in the literature. To aid in comparison, we divide our sample into two redshift bins (\mbox{4.317 $\leq z <$ 5.235} and \mbox{5.236 $\leq z <$ 6.338}) that have approximately equal path lengths and align well with the redshift intervals covered by previous surveys. Our sample has a total absorption path of \mbox{$\Delta X$ = 105.9} in both redshift bins; a factor of $\sim$~2.6 ($\sim$~1.4) larger than the previous largest dataset at $z \gtrsim$~5.2 \mbox{(4.3 $\lesssim z \lesssim$ 5.2)}. This dataset represents the largest homogeneous set of \CIV\ absorbers at $z\gtrsim$~5 measured at \mbox{S/N~$\gtrsim$~10} with spectral resolution \mbox{FWHM $\lesssim$~30~\kms}.

\section{C~IV Line Statistics}\label{sec:results}

\subsection{Number Density}\label{subsec:number_density}
The number density \dndx\ is defined as the number of absorbers per unit comoving path length. We investigate how the number density of \CIV\ absorbers evolves over the probed redshift range by dividing the sample into four redshift bins, listed in Table \ref{table:measurements}. The bins were chosen to divide the survey pathlength approximately equally and provide measurements of \dndx\ with a time resolution of $\sim$~100~Myr, enabling us to robustly constrain the rate of evolution in the number density of \CIV\ absorbers over \mbox{4.3 $\lesssim z \lesssim$ 6.3}.

The measured number density is strongly dependent on the considered column density range. We calculate \dndx\ over the entire column density range of our sample (\mbox{13.2 $\leq \log N <$ 15.0}). To enable completeness corrections to be applied, the absorbers in each redshift bin are divided into column density bins with a width of 0.1 dex (28 bins in total), and \dndx\ is computed as follows:
\begin{equation}
 \frac{dn}{dX}(z) = \sum_{i=1}^{28} \frac{n(\log N_i)}{C(z, \log N_i)}
\end{equation}
Here, $n(\log N_i)$ represents the number of absorbers in the $i$th $\log N$ bin and $C(z,\log N_i)$ is the sample completeness at the relevant $z$ and $\log N$. The latter is computed using the arctan fits provided in \citetalias{Davies22Survey} (see Equation 3 and Table 2 therein) which are shown in the right-hand panel of Figure \ref{fig:completeness}. We adopt the lower redshift completeness curve for all measurements at $z \leq$~5.235 and the higher redshift curve for all measurements at $z >$~5.235. We note that very similar results are obtained when calculating the completeness correction factors directly from the mock absorber catalogs (following the method described in \citetalias{Davies22Survey}) over the exact redshift range of each bin. 

We estimate the errors on \dndx\ both assuming Poisson statistics and using bootstrap resampling. For each redshift bin, the bootstrap error is calculated by selecting quasar sightlines for which the absorption pathlength over that redshift bin is non-zero, randomly sampling $t$ times from this set of $t$ sightlines (with replacement), measuring \dndx, repeating this 200 times, and computing the standard deviation of these measurements. We also report Poisson errors based on the 1$\sigma$ upper and lower bounds on $n$ calculated using Equations 9 and 14 from \citet{Gehrels86}, respectively. Our measurements are listed in Table \ref{table:measurements}. The bootstrap errors are typically larger than the Poisson errors because the bootstrap sampling accounts for factors such as variation in S/N (and therefore completeness) between spectra and cosmic variance. We adopt the boostrap errors in the subsequent analysis.

\begin{figure}
 \includegraphics[scale=0.55, clip = True, trim = 10 0 0 0]{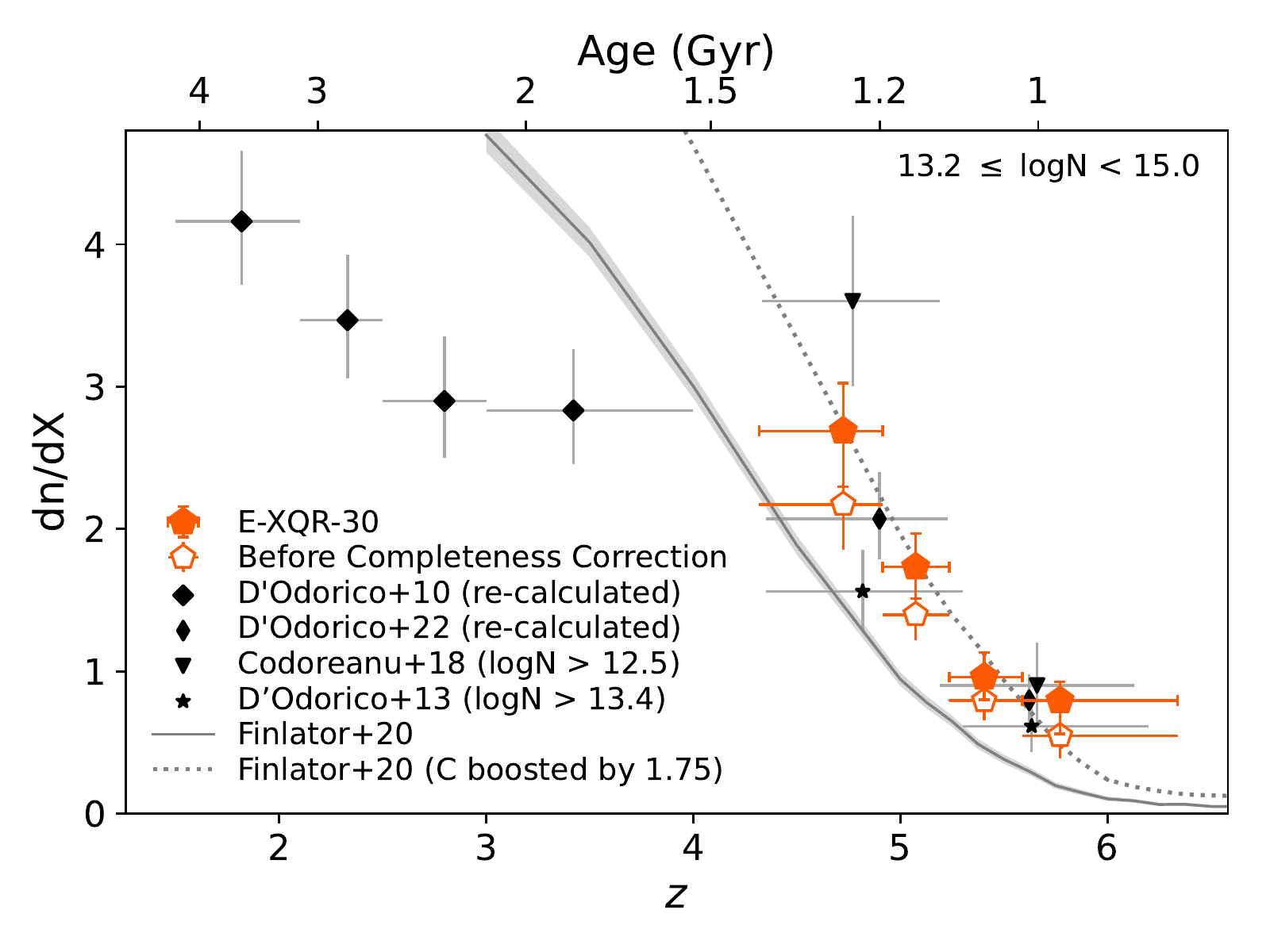} 
\caption{Number density of \CIV\ absorbers over the column density range \mbox{13.2 $\leq \log N <$ 15.0} in four redshift bins (orange markers). The open and solid markers show measurements before and after correcting for completeness, respectively. Horizontal error bars indicate the widths of the redshift bins and vertical error bars are calculated using bootstrapping. Black markers show measurements from the literature with minimum column densities indicated in the legend. The grey curves show predictions from the \citet{Finlator20} cosmological simulation over the same redshift and column density ranges, before (solid) and after (dotted) rescaling the carbon abundances (see Section \ref{subsec:finlator20}).} \label{fig:number_density} 
\end{figure}

The number density measurements are shown in Figure \ref{fig:number_density}. The open and filled orange markers show the values before and after correcting for completeness, respectively, and the black markers show measurements from other studies in the literature. The statistics for the \citet{DOdorico10} and \citet{DOdorico22} samples have been re-calculated to match the $\log N$ ranges used in this paper. The errors on $z$ and \dndx\ indicate the widths of the redshift bins and the bootstrap errors, respectively.  

\begin{table*}
\begin{nscenter}
\begin{tabular}{cccccccccc}
\hline
 & & & &  & $\delta$(dn/dX) & $\delta$(dn/dX) & \OmCIV\ & $\delta$\OmCIV\ $\left(\times 10^{-8} \right)$ & $\delta$\OmCIV\ $\left(\times 10^{-8} \right)$ \\ $z$ range & $<z>$ & $\Delta X$ & \# & dn/dX & (Bootstrap) & (Poisson) & $\left(\times 10^{-8}\right)$ & (Bootstrap) & (SL96) \\ \hline \multicolumn{10}{c}{13.2 $\leq \log (N/{\rm cm}^2) <$ 15.0} \\ \hline 
4.317 $\leq z <$ 4.915 & 4.72 & 52.93 & 115 & 2.69 & 0.39, 0.33 & 0.25, 0.27 & 2.63 & 0.66, 0.59 & 0.43 \\ 
4.915 $\leq z <$ 5.236 & 5.07 & 52.94 & 74 & 1.73 & 0.22, 0.23 & 0.20, 0.23 & 1.37 & 0.27, 0.28 & 0.19 \\ 
5.236 $\leq z <$ 5.590 & 5.40 & 52.94 & 42 & 0.96 & 0.16, 0.17 & 0.15, 0.17 & 0.94 & 0.16, 0.21 & 0.18 \\ 
5.590 $\leq z <$ 6.339 & 5.77 & 52.96 & 29 & 0.79 & 0.23, 0.13 & 0.15, 0.18 & 0.55 & 0.22, 0.16 & 0.14 \\ \hline\multicolumn{10}{c}{13.8 $\leq \log (N/{\rm cm}^2) <$ 15.0} \\ \hline 
4.317 $\leq z <$ 4.915 & 4.72 & 52.93 & 24 & 0.48 & 0.19, 0.14 & 0.10, 0.12 & 1.52 & 0.63, 0.57 & 0.42 \\ 
4.915 $\leq z <$ 5.236 & 5.07 & 52.94 & 13 & 0.26 & 0.08, 0.09 & 0.07, 0.09 & 0.67 & 0.21, 0.26 & 0.17 \\ 
5.236 $\leq z <$ 5.590 & 5.40 & 52.94 & 12 & 0.24 & 0.06, 0.07 & 0.07, 0.09 & 0.56 & 0.17, 0.18 & 0.17 \\ 
5.590 $\leq z <$ 6.339 & 5.77 & 52.96 & 8 & 0.16 & 0.08, 0.06 & 0.06, 0.08 & 0.32 & 0.19, 0.14 & 0.14 \\ \hline
\end{tabular}
\end{nscenter}
\caption{Completeness-corrected \CIV\ absorber statistics measured after grouping components with $\Delta v <$~50~\kms. The statistics are reported in two column density ranges and four redshift bins chosen to cover approximately equal absorption path lengths. The \OmCIV\ values given in this table are calculated using Equation \ref{eqn:omciv}. We provide two sets of error estimates for each of \dndx\ and \OmCIV. The \citetalias{StorrieLombardi96} approximation for $\delta$\OmCIV\ is calculated using Equation \ref{eqn:civ_err}. In each error column, the two listed values represent the lower and upper 1$\sigma$ error.}\label{table:measurements}
\end{table*}

We find that the number density of \CIV\ absorbers decreases by a factor of 3.4~$\pm$~0.9 over the $\sim$~300~Myr interval between $z\sim$~4.7 and $z\sim$~5.8. Interestingly, $\sim$~91\% of this drop occurs over the 200~Myr interval between $z\sim$~4.7 and $z\sim$~5.4. Our results are qualitatively consistent with the findings of \citet{DOdorico22} who report a sharp drop in the \CIV\ number density between $z\sim$~4.7 and $z\sim$~5.3, and \citet{Codoreanu18} who measured a significant decrease in the \CIV\ number density between $z\sim$~4.8 and $z\sim$~5.7. 

The large number of \CIV\ absorbers used in this study enables us to robustly confirm that the decline in \CIV\ absorption over \mbox{4.3 $\lesssim z \lesssim$ 6.3} is much more rapid than what is observed over \mbox{1.5 $< z <$ 4}. A factor of 3.4 decrease in \dndx\ over a period of 300 Myr corresponds to approximately a factor of 1.5 decrease in \dndx\ every 100 Myr. A similar decrease in number density (a factor of 1.5~$\pm$~0.3) is measured over \mbox{1.8 $\lesssim z \lesssim$ 3.4} from the \citet{DOdorico10} sample, but this covers an order of magnitude more cosmic time \mbox{(1.75 Gyr)} than the E-XQR-30 sample. The number density we measure in the $z\sim$~4.7 bin is very similar to the value measured at $z\sim$~3.4, suggesting that the rapid decline in \dndx\ likely commences at $z\simeq$~5. Our results clearly show that the number density of \CIV\ absorbers decreases slowly over \mbox{1 $\lesssim z \lesssim$ 5} before declining rapidly towards higher redshifts.

\subsection{Column Density Distribution Function (CDDF)}\label{subsec:cddf}
The column density distribution function (CDDF) is a fundamental property of an absorber population, similar to the galaxy luminosity function. It is defined as the number of absorption systems per unit column density per unit redshift absorption path d$X$, and is denoted as $f(N)$. A population of absorbers with constant distributions of physical cross-sections and co-moving space densities illuminated by a non-evolving ionizing spectrum will have the same CDDF at all redshifts.

Measuring the CDDF at different redshifts is important because it can reveal whether the observed drop in \CIV\ number density is dominated by absorbers in a particular column density range. If only the weakest (strongest) absorbers are suppressed at higher redshifts, then the CDDF would be flatter (steeper) at earlier times. To accurately constrain the slope of the CDDF we require as many absorbers per redshift interval as possible. Therefore, we characterize the CDDF for the two redshift intervals shown in Figure \ref{fig:path_length_comparison} (centered at $z\sim$~4.9 and $z\sim$~5.6). 

\begin{figure*}
 \includegraphics[scale=0.55]{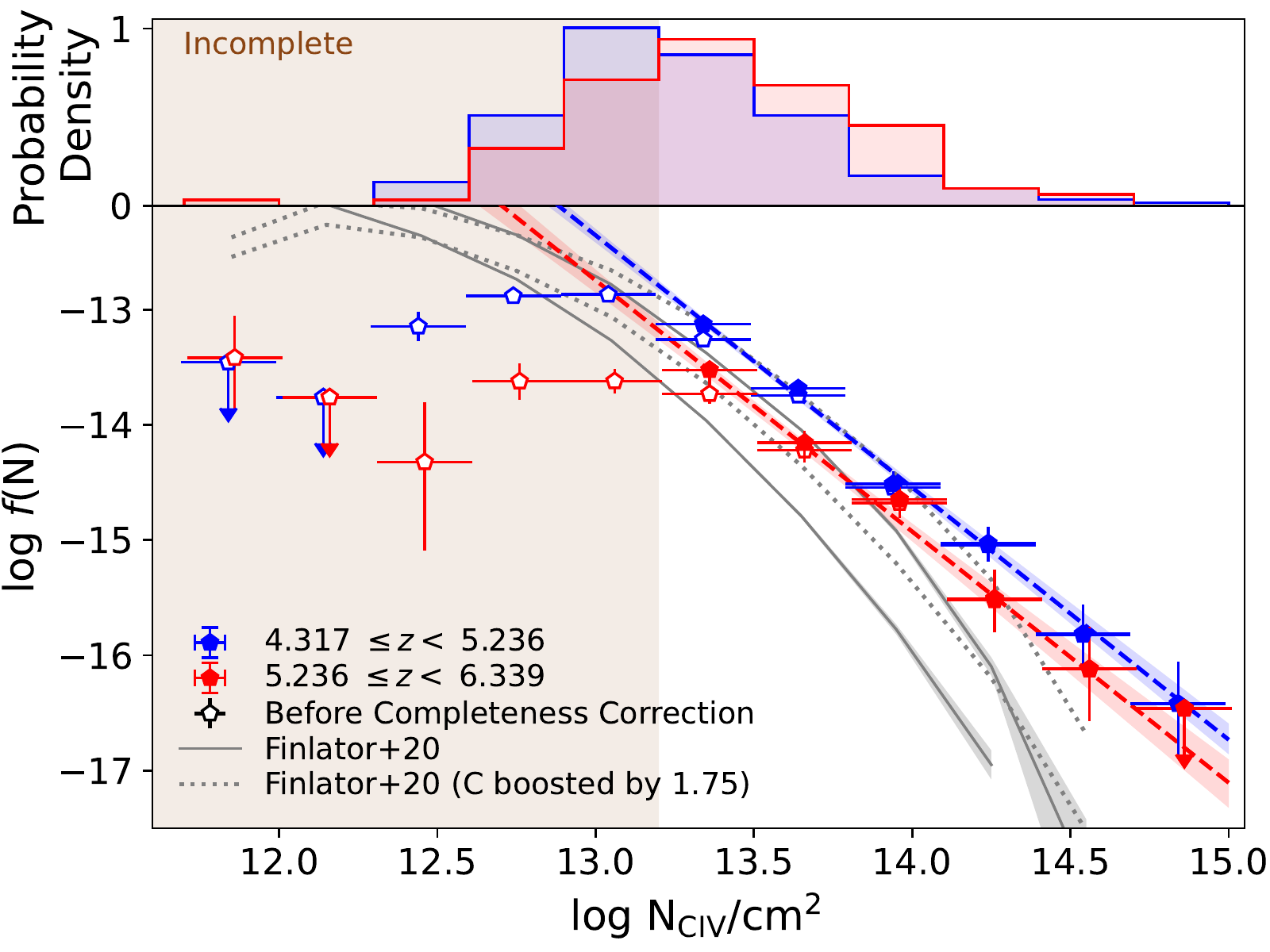}\includegraphics[scale=0.55]{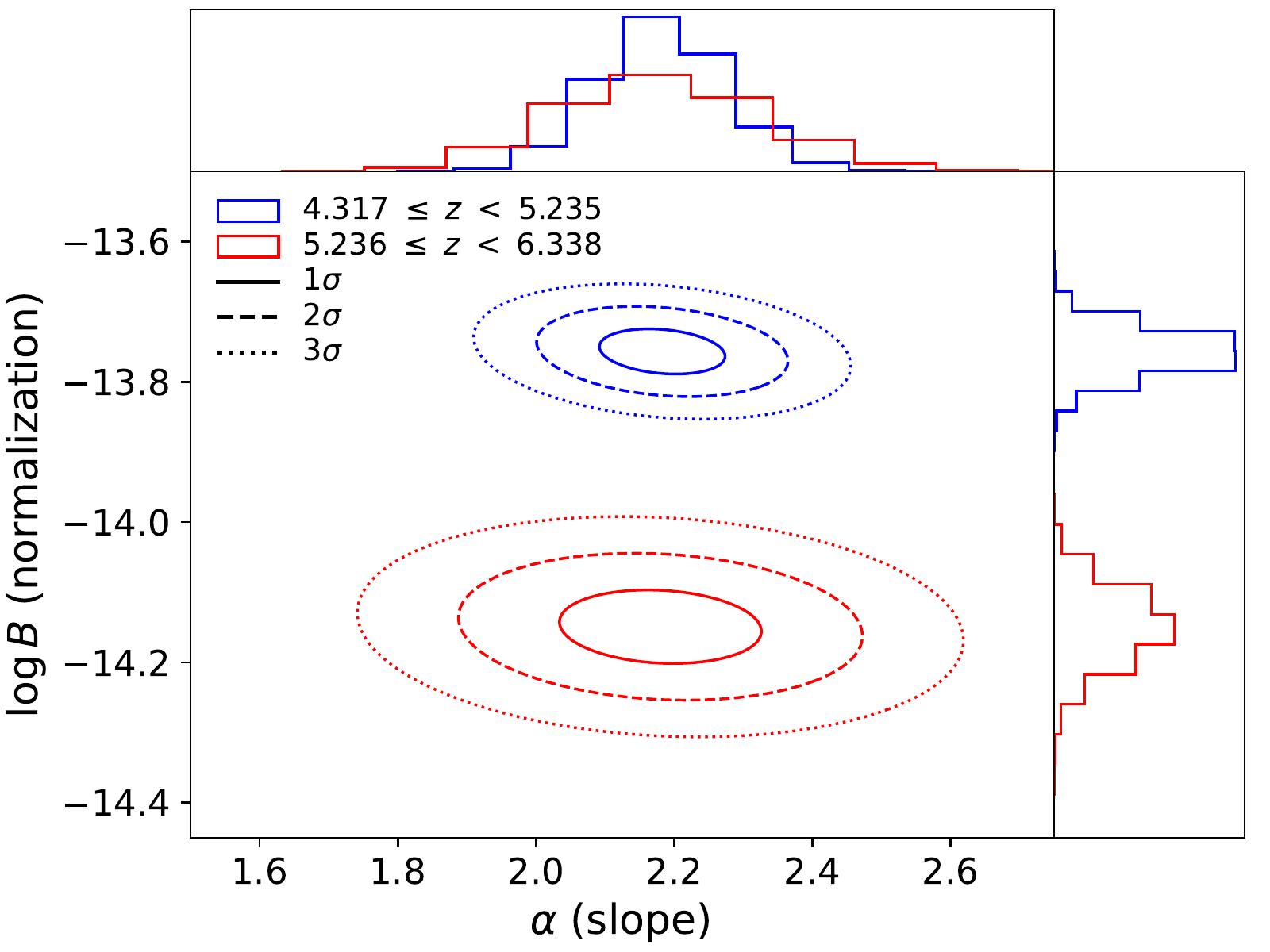}
 \caption{Left: The \CIV\ Column Density Distribution Function (CDDF) in two redshift bins centered at $z\sim$~4.9 (blue) and $z\sim$~5.6 (red). Within each redshift bin, the absorbers are split into 11 $\log N$ bins each with a width of 0.3 dex spanning \mbox{11.7 $< \log N <$ 15.0}. The open and solid markers show measurements before and after correcting for completeness, respectively. We do not  compute completeness-corrected values in the range where the survey is incomplete (\mbox{$\log N <$ 13.2}, brown shaded region). Vertical error bars represent the uncertainties assuming Poisson statistics. Upper limits are reported for bins where no absorbers are detected. We use a maximum likelihood approach to fit a power law distribution to the unbinned absorber data at each redshift. The red and blue dashed lines and shaded regions show the best fit and the associated 68\% confidence interval at each redshift, respectively. Grey curves show the CDDFs of simulated absorbers from the \citet{Finlator20} dataset over the same two redshift ranges as our measurements. Right: 1$\sigma$ (solid), 2$\sigma$ (dashed), and 3$\sigma$ (dotted) confidence intervals of the joint posterior probability distribution functions for $\alpha$ (slope) and $B$ (normalization) for both redshift intervals. We measure a strong decrease in the normalization of the CDDF (0.38 dex) but no significant evolution in the slope, demonstrating that the decline in \CIV\ absorption occurs approximately uniformly across \mbox{13.2 $\leq \log N <$ 15.0}.} \label{fig:cddf} 
\end{figure*}

To visualize the CDDF at each redshift, we first measure $f(N)$ without completeness corrections over the whole column density range covered by the sample of primary intervening \CIV\ absorbers (\mbox{11.7 $< \log N <$ 15.0}). Within each redshift bin, the absorbers are split into 11 $\log N$ bins with a width of 0.3~dex. We also measure completeness-corrected values for bins lying above the 50\% completeness limit \mbox{($\log N \geq$ 13.2)} by further splitting each $\log N$ bin into three sub-bins of width 0.1 dex and computing $f(N)$ as follows:
\begin{equation}
 f(N_i,z) = \log_{\rm 10} \left(\frac{1}{\Delta N_i \Delta X} \sum_{j=1}^{3} \frac{n(\log N_{i,j})}{C(z, \log N_{i,j})} \right)
\end{equation}

Here, $\Delta N_i$ is the column density range (in linear space) covered by the $i$th $\log N$ bin. When no absorbers are detected in a given bin, we report an upper limit based on the Poisson upper bound for $n$ \citep{Gehrels86}. 

The $f(N)$ measurements are shown in the left-hand panel of Figure \ref{fig:cddf} and listed in Table \ref{cddf_table}. The values before and after correcting for completeness are shown as open and solid markers, respectively. These binned measurements are not used when fitting the CDDF (as explained below), so we report only Poisson errors. The CDDF rises in the region where the sample is incomplete (brown shaded region) before declining towards larger column densities. As expected, the completeness corrections have a relatively minor impact on $f(N)$ for bins above the completeness limit.

\begin{table}
\begin{nscenter}
\begin{tabular}{lccc}
\hline
 $\log$N range & \# & $\log f(N)$ & $\delta\log f(N)$ (Poisson) \\ \hline
\multicolumn{4}{c}{4.317 $\leq z <$ 5.235 ($<z>$ = 4.914)} \\ \hline
13.20 -- 13.50 & 92 & -13.13 & 0.06,0.04 \\ 
13.50 -- 13.80 & 60 & -13.69 & 0.06,0.06 \\ 
13.80 -- 14.10 & 19 & -14.51 & 0.11,0.11 \\ 
14.10 -- 14.40 & 12 & -15.03 & 0.15,0.14 \\ 
14.40 -- 14.70 & 4 & -15.81 & 0.28,0.25 \\ 
14.70 -- 15.00 & 2 & -16.42 & 0.45,0.36 \\ 
\hline
\multicolumn{4}{c}{5.236 $\leq z <$ 6.338 ($<z>$ = 5.589)} \\ \hline
13.20 -- 13.50 & 31 & -13.52 & 0.12,0.05 \\ 
13.50 -- 13.80 & 20 & -14.16 & 0.11,0.10 \\ 
13.80 -- 14.10 & 14 & -14.65 & 0.13,0.13 \\ 
14.10 -- 14.40 & 4 & -15.51 & 0.28,0.25 \\ 
14.40 -- 14.70 & 2 & -16.12 & 0.45,0.36 \\ 
14.70 -- 15.00 & 0 & $<$ -16.46 & -- \\ 
\hline
\end{tabular}
\caption{Completeness-corrected measurements of $f(N)$ in two redshift intervals. Within each redshift interval, the absorbers are split into $\log N$ bins of width 0.3~dex. When no absorbers are detected we report an upper limit based on the Poisson upper bound on the number of absorbers.}\label{cddf_table}
\end{nscenter}
\end{table}

\begin{table*}
\begin{nscenter}
\begin{tabular}{cccccc}
\hline
 &  & & & \multicolumn{2}{c}{\OmCIV$\left(\times 10^{-8}\right)$} \\
$z$ range & $<z>$ & $\log B$ & $\alpha$ & 13.2 $\leq \log N/{\rm cm}^2 <$ 15.0 & 13.8 $\leq \log N/{\rm cm}^2 <$ 15.0 \\ \hline
4.317 $\leq z <$ 5.236 & 4.914 & -13.76$^{+0.03}_{-0.03}$ & 2.19$^{+0.08}_{-0.10}$ & 1.99$^{+0.24}_{-0.21}$ & 1.15$^{+0.22}_{-0.19}$ \\ 
5.236 $\leq z <$ 6.339 & 5.589 & -14.14$^{+0.04}_{-0.06}$ & 2.18$^{+0.15}_{-0.15}$ & 0.81$^{+0.16}_{-0.13}$ & 0.47$^{+0.15}_{-0.12}$ \\ \hline
\end{tabular}
\end{nscenter}
\caption{Parameters of the power law distributions (Equation \ref{eqn:powerlaw}) fit to the unbinned CIV absorber statistics using the maximum likelihood approach described in Section \ref{subsec:cddf}, and \OmCIV\ values computed from the integral of the CDDF (Equation \ref{eqn:cddf_integral}). The errors on $\alpha$ and $\beta$ represent the 1$\sigma$ confidence intervals measured from the posterior probability distribution functions shown in the right-hand panel of Figure \ref{fig:cddf}. The errors on \OmCIV\ represent the 68\% confidence interval of values obtained by calculating \OmCIV\ for every realization of $\alpha$ and $\beta$.}\label{cddf_fit_table}
\end{table*}

The binned measurements help to visualize the slope and normalization of the CDDF in the two redshift bins. However, to minimize biases associated with binning, we fit the CDDF using the binning-independent maximum likelihood approach outlined in \citet{Bosman17}. We parametrize the CDDF as a power law distribution with the form 
\begin{equation}\label{eqn:powerlaw}
f(N) = B \left(\frac{N}{N_0}\right)^{-\alpha}
\end{equation}
We adopt \mbox{$N_0 = 10^{13.64}$} for consistency with previous works \citep{DOdorico13, Codoreanu18}, but note that our conclusions remain the same regardless of the adopted value. The likehood function is written as follows:
\begin{equation}
 \mathcal{L}(B,\alpha) = P_n(n|B,\alpha) \, \, \prod_i P_i(N_i|\alpha)
\end{equation}
The first term $P_n(n|B,\alpha)$ represents the Poisson probability of observing a total of $n$ absorbers across \mbox{13.2 $\leq \log N <$ 15.0} given a CDDF with parameters \mbox{($B$, $\alpha$)}, for which the expected number of absorbers is given by
\begin{equation}\label{eqn:nexpected}
 n_{\rm expected}(B,\alpha) = B \Delta X \times 10^{N_0 \alpha} \int_{10^{13.2}}^{10^{15.0}} \frac{C(z, N)}{N^\alpha} dN
\end{equation}
The second term $P_i(N_i|\alpha)$ represents the Poisson probability of observing a system with column density $N_i$ when $B$ is scaled such that \mbox{$P_n(n|B,\alpha)$ = 1}. We calculate this probability by comparing the number of observed and expected absorbers in bins of width 0.1 dex, because the mean uncertainty on the individual $\log N$ measurements is 0.09~dex. However, we note that adopting a bin size of 0.01~dex produces consistent results.

We perform the fitting using the Affine Invariant Markov Chain Monte Carlo (MCMC) Ensemble Sampler \textsc{emcee} \citep{ForemanMackey13}. We adopt uniform priors on $\alpha$ and $B$ over the ranges \mbox{$-20 < \log_{\rm 10} B < -10$} and \mbox{0 $\leq \alpha \leq$ 4}. We run the MCMC with 300 walkers, 100 burn-in steps, and 500 run steps. The initial parameter estimates for the walkers are distributed uniformly over the ranges \mbox{$-$13 $\leq \log B_{\rm 10} \leq$ $-$17} and \mbox{0 $\leq \alpha \leq$ 4}. 

The right-hand panel of Figure \ref{fig:cddf} shows the 1$\sigma$, 2$\sigma$ and 3$\sigma$ contours of the joint posterior probability distribution function for $\alpha$ and $B$, and the best-fit values and 1$\sigma$ errors on both parameters are listed in Table \ref{cddf_fit_table}. The dashed lines and shaded regions in the left-hand panel of Figure \ref{fig:cddf} show the best-fit power law distributions and the 68\% confidence intervals, respectively.

We find that the normalization of the CDDF ($B$) decreases strongly over cosmic time, changing by \mbox{$\Delta B$ = $-$0.38~$\pm$~0.06~dex} (a factor of 2.4~$\pm$~0.3) between $z\sim$~4.9 and $z\sim$~5.6. This is consistent with the observed drop in number density over the same redshift interval. Our findings are in agreement with previous studies which report consistent decreases in normalization over a similar redshift range (\citealt{Codoreanu18} measure \mbox{$\Delta B$ = $-$0.47~$\pm$~0.20 dex} and \citealt{DOdorico13} report a factor of $\sim$~2~--~3 decrease in normalization). However, the large size of the E-XQR-30 sample enables us to measure this evolution at a significance of $>$~5$\sigma$ for the first time. The \citet{Codoreanu18} sample includes 41 systems over 4 sightlines and the \citet{DOdorico13} sample includes 102 systems over 6 sightlines at \mbox{4.35 $< z <$ 6.2}, compared to the current sample of 260 systems over 42 sightlines.

In contrast, there is no evidence for evolution in the slope of the CDDF over \mbox{4.3 $\lesssim z \lesssim$ 6.3}. We measure power law slopes of \mbox{$\alpha$ = 2.19$^{+0.08}_{-0.10}$} at $z\sim$~4.9 and \mbox{$\alpha$ = 2.18~$\pm$~0.15} at $z\sim$~5.6, corresponding to a change in slope of \mbox{$\Delta \alpha$ = $-$0.01~$\pm$~0.17}. This is consistent with previous works which also found $\Delta \alpha \simeq$~0 over a similar redshift range (e.g. $\Delta \alpha$~=~$-$0.18~$\pm$~0.36; \citealt{DOdorico13}, $\Delta \alpha$~=~0.47~$\pm$~0.47; \citealt{Codoreanu18}). However, our measurements provide a more stringent constraint on $\Delta \alpha$ due to the significant increase in sample size.

We note that the individual $\alpha$ values measured in this work are larger than the values reported by \citet{DOdorico13} and \citet{Codoreanu18} which range from 1.44 to 1.96. This likely stems from a combination of factors. Firstly, there are some differences in adopted methodology between the studies. \citet{Codoreanu18} group components with $\Delta v <$~500~\kms\ which is expected to result in a flatter CDDF than the 50~\kms\ grouping used in this work \citep[e.g.][]{Boksenberg15}. If we group components with $\Delta v <$~500~\kms, we measure \mbox{$\alpha \simeq$ 1.8~--~2.0}, in better agreement with the values published by \citet{Codoreanu18}. \citet{DOdorico13} performed least squared fitting on the binned measurements in contrast to the unbinned maximum likelihood approach adopted in this work. Finally, the samples used in the previous studies had similar completeness limits to our sample but very few \CIV\ absorbers with \mbox{$\log N >$~13.8} (14 in \citealt{DOdorico13} and 8 in \citealt{Codoreanu18} over both redshift intervals combined). The current dataset includes 54 such systems, increasing the dynamic range in column density over which reliable measurements can be made and thereby improving the constraint on $\alpha$. 

Put together, our results provide the first robust indication that the observed decrease in the \CIV\ \dndx\ is not dominated by absorbers in a particular column density range but occurs approximately uniformly for absorbers spanning \mbox{13.2 $\leq \log N <$ 15.0}. 

\begin{figure*}
 \includegraphics[scale=0.55]{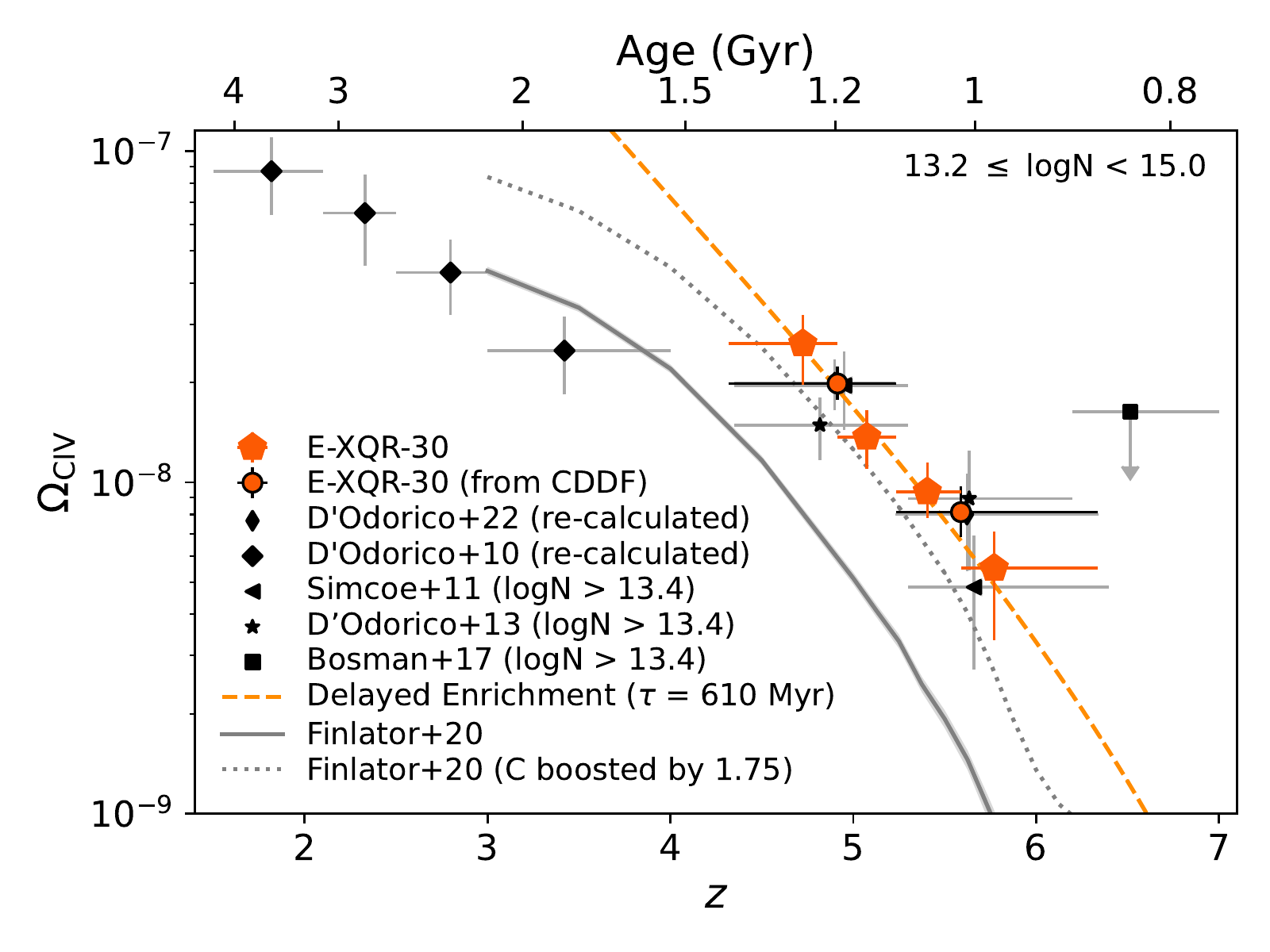}\includegraphics[scale=0.55]{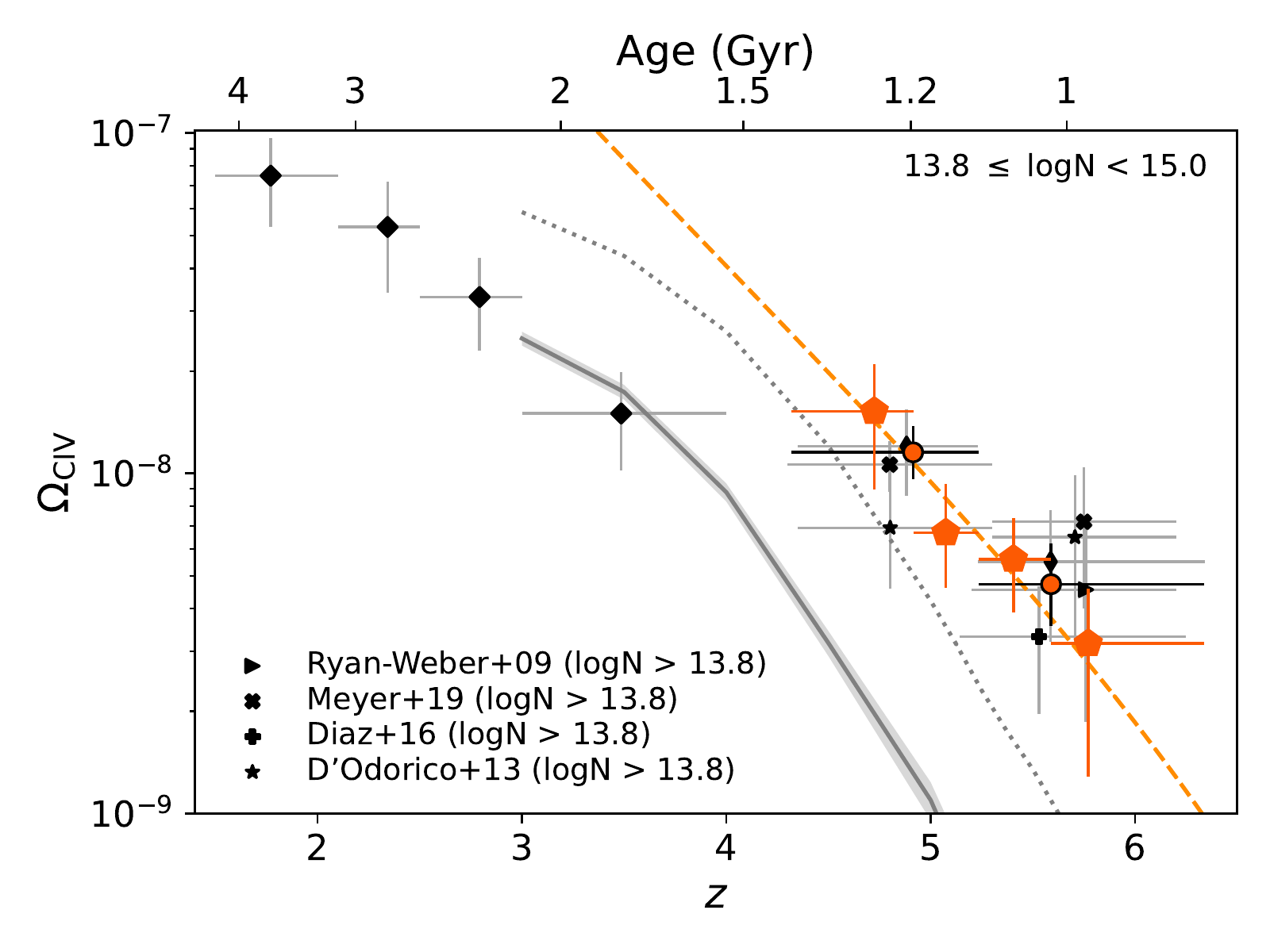} \\
\caption{Redshift evolution of \OmCIV\ measured over \mbox{13.2 $\leq \log N <$ 15.0} (left) and \mbox{13.8 $\leq \log N <$ 15.0} (right). The orange markers with and without black borders represent measurements made using the integral of the CDDF (Equation \ref{eqn:cddf_integral}) and the \citetalias{StorrieLombardi96} approximation (Equation \ref{eqn:omciv}), respectively. Orange dashed curves show the best-fit delayed enrichment model for each set of measurements (for the \citealt{Madau14} fit to the cosmic star formation rate density, as described in Section \ref{subsec:delayed_enrichment}). Other plotting symbols and errors are the same as in Figure \ref{fig:number_density}. Literature measurements of \OmCIV\ have been scaled to match our adopted cosmology. The large \mbox{E-XQR-30} dataset enables us to robustly measure \OmCIV\ at intervals of $\sim$~100~Myr, revealing a smooth but steep decline from $z\sim$~4.7 to $z\sim$~5.8.} \label{fig:omciv} 
\end{figure*} 

\subsection{Cosmic Mass Density (\OmCIV)}\label{subsec:om_civ}
The \CIV\ cosmic mass density ($\Omega_{\rm C IV}$) is defined as the \CIV\ mass per unit comoving Mpc, $\rho_{\rm{C}\textsc{IV}}$, divided by the critical density of the Universe at $z\sim$~0, $\rho_{\rm crit}$. The comoving mass density of \CIV\ can be obtained by integrating the CDDF: 
\begin{equation}\label{eqn:cddf_integral}
 \rho_{\rm{C}\textsc{IV}} = \frac{H_0 m_{\rm C}}{c} \int N f(N) dN  
\end{equation}
where $H_0$ is the Hubble constant and $m_C$ is the atomic mass of carbon. The \CIV\ cosmic mass density is then given by:
\begin{equation}
 \Omega_{\rm C \textsc{IV}} = \rho_{\rm{C}\textsc{IV}} / \rho_{\rm crit}
\end{equation}
\citet[][hereafter \citetalias{StorrieLombardi96}]{StorrieLombardi96} proposed that the integral of the CDDF can be approximated as $\Sigma N/\Delta X$, yielding the following expression:
\begin{equation}\label{eqn:pciv_approx}
 \Omega_{\rm{C}\textsc{IV}} \simeq \frac{H_0 m_{\rm C}}{c \rho_{\rm crit}} \frac{\Sigma N}{\Delta X}  
\end{equation}
Many previous studies of \CIV\ absorption at \mbox{$z\sim$~5~--~6} have used this approximation to measure \OmCIV\ due to small sample sizes \citep[e.g.][]{RyanWeber09, Diaz16, DOdorico13, Meyer19, DOdorico22}. We present measurements of \OmCIV\ obtained using both the integral of the CDDF and the \citetalias{StorrieLombardi96} approximation.

The measured \OmCIV\ is strongly dependent on the adopted $\log N$ range. Many previous studies report \CIV\ absorber statistics over \mbox{13.8 $\leq \log N <$ 15.0}, but our sample is $\geq$~50\% complete all the way down to $\log N$~=~13.2. Therefore, we present \OmCIV\ measurements summed over both \mbox{13.2 $\leq \log N <$ 15.0} and \mbox{13.8 $\leq \log N <$ 15.0}.

The \OmCIV\ values calculated from the integral of the CDDF are already completeness-corrected because the completeness is taken into account when fitting the power law distribution (see Equation \ref{eqn:nexpected}). The errors on these measurements are computed by calculating \OmCIV\ for each ($B$, $\alpha$) pair in the MCMC joint posterior distribution and taking the standard deviation of these values. The measurements obtained from Equation \ref{eqn:cddf_integral} and the corresponding MCMC errors are listed in Table \ref{cddf_fit_table}.

We use the \citetalias{StorrieLombardi96} approximation to compute \OmCIV\ for the four redshift bins listed in Table \ref{table:measurements}. Once again, to enable completeness corrections, the absorbers in each redshift bin are divided into column density bins with a width of 0.1 dex and \OmCIV\ is calculated as follows:
\begin{equation}\label{eqn:omciv}
\Omega_{\rm{C}\textsc{IV}} \simeq \frac{H_0 m_{\rm C}}{c \rho_{\rm crit} \Delta X} \sum_{i} \frac{N_{{\rm total},i}}{C(z, \log N_i)}  
\end{equation}
Here, $N_{{\rm total},i}$ is the total column density of all absorbers in the $i$th $\log N$ bin. The uncertainty on \OmCIV\ is typically computed using the following approximation for the fractional variance, also proposed by \citetalias{StorrieLombardi96}:
\begin{equation}\label{eqn:civ_err}
 \left(\frac{\delta \Omega_{\rm{C} \textsc{IV}}}{\Omega_{\rm{C} \textsc{IV}}}\right)^2 = \frac{\Sigma \left(N^2\right)}{\left(\Sigma N\right)^2}
\end{equation}
This approximation is widely adopted in the literature but is known to underestimate the true errors by up to a factor of 1.5 \citep[e.g.][]{DOdorico10}. As previously discussed, bootstrapping provides a more accurate estimate of the errors for sufficiently large samples because it accounts for factors such as variations in completeness between spectra. Table \ref{table:measurements} lists the \OmCIV\ measurements calculated using Equation \ref{eqn:omciv} along with errors calculated using both the \citetalias{StorrieLombardi96} approximation (Equation \ref{eqn:civ_err}) and bootstrapping. We adopt the bootstrap errors in our analysis.

The results are shown in Figure \ref{fig:omciv}. The left-hand and right-hand panels show \OmCIV\ measurements summed over \mbox{13.2 $\leq \log N <$ 15.0} and \mbox{13.8 $\leq \log N <$ 15.0}, respectively. The measurements in the left-hand panel are based on a larger number of absorbers (see Table \ref{table:measurements}) and therefore have smaller associated errors than those in the right-hand panel. The orange markers with and without black borders show the measurements computed from the integral of the CDDF (Equation \ref{eqn:cddf_integral}) and the \citetalias{StorrieLombardi96} approximation (Equation \ref{eqn:omciv}), respectively. The two methods produce consistent results. In each panel, the black markers show literature measurements for which the minimum $\log N$ is consistent with our adopted value to within 0.2 dex. The literature measurements have been scaled to match our adopted cosmology. 

Our results are consistent with previously published values, but the larger sample size enables us to measure \OmCIV\ in a larger number of redshift bins and therefore track the time evolution more finely than previous studies. We find that \OmCIV\ declines steeply but smoothly between $z\sim$~4.7 and $z\sim$~5.8, decreasing by a factor of 4.8~$\pm$~2.0 over this interval of only 300 Myr for absorbers with \mbox{13.2 $\leq \log N <$ 15.0}. (The values measured over \mbox{13.8 $\leq \log N <$ 15.0} yield consistent results but at a lower statistical significance due to the smaller sample size). 

Combining our results with measurements at \mbox{1.5 $< z <$ 4} from \citet{DOdorico10}, we find that \OmCIV\ evolves much more rapidly at $z\gtrsim$~4.3 than at lower redshifts. \OmCIV\ decreases by a factor of 3.5~$\pm$~1.3 over the 1.75 Gyr interval from $z\sim$~1.8 to $z\sim$~3.4; comparable to the decline between $z\sim$~4.7 and $z\sim$~5.8 which occurs over a much shorter time interval. The \OmCIV\ values measured at $z\sim$~4.7 and $z\sim$~3.4 are consistent with one another, suggesting that, similar to \dndx, the rapid decline in \OmCIV\ commences at $z\gtrsim$~5.

Our results are consistent with many previous works which also suggested that \OmCIV\ remains approximately constant over \mbox{1.5 $\lesssim z \lesssim$ 5} before declining rapidly towards higher redshifts \citep[e.g.][]{Songaila97, Songaila01, Songaila05, Becker09, RyanWeber09, DOdorico10, Simcoe11, DOdorico13, Boksenberg15, Diaz16, Bosman17, Codoreanu18, Meyer19, DOdorico22}. However, the improved measurements presented in this paper constrain the rate and timing of the accelerated decline in \OmCIV\ to much higher statistical significance than was previously possible. 

\subsection{Comparison to Predictions from the Technicolor Dawn Simulation}\label{subsec:finlator20}
We compare our results to predictions from the \textit{Technicolor Dawn} cosmological simulation (described in detail in \citealt{Finlator20}). The simulation is run using a custom version of \textsc{Gadget-3} and covers a volume of 15$h^{-1}$ Mpc at a mass resolution of 2.6~$\times$~10$^5 M_\odot$. Metal enrichment is modelled using the supernova yields of \citet{Nomoto06} scaled to the assumed \citet{Kroupa01} IMF with a hypernova fraction of 50\%. Stellar feedback is implemented using momentum kicks tuned to match the \citet{Muratov15} outflow mass-loading factor scaling, but with a slight reduction in normalization to better match the observed galaxy stellar mass and UV luminosity functions. The outflow velocities are taken from \citet{Muratov15} with two boost factors applied to produce larger outflow velocities that are more consistent with observed values \citep[see also][]{Dave16}. The quasar radiation field is assumed to be spatially uniform while the inhomogeneous galaxy UV background is modelled by solving the multi-frequency radiative transfer equation for galaxies on a Cartesian grid. A mock absorber catalogue was generated by passing sightlines through the simulation volume. The metallicity of the gas particles is traced self-consisently in the simulation, while the ionization state is computed using ionization equilibrium calculations \citep[see][]{Finlator15}. 

The mock absorber catalog has previously been compared with measured absorber statistics at \mbox{3 $\lesssim z \lesssim$ 5} \citep{Hasan20} and \mbox{5 $\lesssim z \lesssim$ 6.3} \citep{Finlator20}. At \mbox{3 $\lesssim z \lesssim$ 5}, \textit{Technicolor Dawn} roughly reproduces the observed statistics of \CIV\ absorbers with $W >$~0.3\AA\ but significantly over-produces weaker absorbers, perhaps indicating that the simulated outflows transport metals too far from galaxies \citep{Hasan20}. At $z\gtrsim$~5, the simulation is able to reproduce the observed CDDF of \SiIV\ absorbers but produces a factor of $\sim$~3 too few \CIV\ absorbers at all column densities \citep{Finlator20}. This tension is also seen in other cosmological simulations \citep[e.g.][]{Rahmati16, Keating16}. \citet{Finlator20} argue that this discrepancy cannot be solved by increasing the strength of the UV background, the star formation efficiency or the overall CGM metallicity because all of these would lead to discrepancies in other galaxy or absorber properties. One potential avenue to reduce the disrepancy could be to relax the assumption that \HII\ regions are ionization bounded. If a non-negligible fraction of \HII\ regions are density bounded, then Lyman continuum photons would be able to escape through attenuated channels, hardening the emergent ionizing spectrum. 

An alternative solution could be to artificially increase the yield of carbon relative to silicon. The carbon yields are impacted by uncertainties on the supernova yields and the initial mass function, which may be more top-heavy in early low-metallicity galaxies \citep[e.g.][]{Nomoto06, Kulkarni13}. \citet{Finlator20} measured mean \CII/\SiII\ rest-frame equivalent width ratios of 0.54 (0.77) for simulated $z\sim$~6 absorbers with \SiII~$\lambda$1260 (\CII~$\lambda$1334) rest-frame equivalent widths exceeding 0.05\AA. These values are 22~--~75\% lower than the mean ratio of 0.94 measured for observed systems at \mbox{5.75 $< z <$ 6.25} in the \citet{Becker19} sample. The mean \CII/\SiII\ equivalent width ratio for absorbers at \mbox{5.75 $< z <$ 6.4} in the \mbox{E-XQR-30} catalog \citepalias{Davies22Survey} is 1.08, supporting the hypothesis that the relative carbon abundances of the simulated absorbers in the \citet{Finlator20} sample could be underestimated. \citet{Finlator20} found that increasing the carbon abundances of the simulated absorbers by a factor of 1.75 brings their number densities into agreement with the observed statistics. 

We compare the \CIV\ absorber statistics predicted by the \citet{Finlator20} simulations to the observed statistics in Figures \ref{fig:number_density}, \ref{fig:cddf} and \ref{fig:omciv}. The solid and dotted grey curves represent the predictions before and after rescaling the carbon abundances by a factor of 1.75, respectively. The curves were calculated using simulated absorbers in the same column density intervals and redshift ranges as the observed systems. Prior to rescaling the carbon abundances, the simulation under-produces \CIV\ absorbers at all column densities. After rescaling the carbon abundances, the predicted number densities are in good agreement with our measurements (Figure \ref{fig:number_density}). The rescaled CDDFs are marginally consistent with the measured statistics near the completeness limit of our sample (\mbox{$\log N \simeq$ 13.2}), but stronger systems are increasingly under-produced moving towards higher column densities (Figure \ref{fig:cddf}). \citet{Hasan20} similarly find that the simulation under-produces strong \CIV\ absorbers at \mbox{3 $\lesssim z \lesssim$ 5}, perhaps due to the limited simulation volume or the absence of quasar-driven spatial fluctuations in the hard UV background. As a direct result, the rescaled \OmCIV\ predictions fall a factor of 1.2~--~2.0 (0.9~--~1.9$\sigma$) below the observed \OmCIV\ computed over \mbox{13.2 $\leq \log N <$ 15.0} and a factor of 1.8~--~4.8 (1.1~--~2.3$\sigma$) below the observed \OmCIV\ computed over \mbox{13.8 $\leq \log N <$ 15.0} (Figure \ref{fig:omciv}). We note that boosting the carbon abundances leads to the over-prediction of \OmCIV\ at $z <$~4, whereas the original predictions are in agreement with the measured values. 

Overall, the discrepancy between the measured and predicted normalization of the \CIV\ CDDF supports the hypothesis that the \citet{Finlator20} simulation underestimates the carbon yield of early stellar populations. However, our robust measurements of strong absorber statistics enabled by the large E-XQR-30 sample size additionaly reveal that the simulation assembles strong systems too slowly.

\section{Discussion}\label{sec:discussion}
The statistical power of the combined sample of 42 quasars considered in this work has enabled us to robustly characterize the evolution of \dndx, \OmCIV, and the slope and normalization of the \CIV\ CDDF over \mbox{4.3 $\lesssim z \lesssim$ 6.3}. Our results conclusively show that the incidence of \CIV\ absorbers decreases rapidly over this redshift range. We additionally show that the slope of the \CIV\ column density distribution function does not evolve significantly \mbox{($\Delta \alpha$ = $-$0.01 $\pm$ 0.17)}, suggesting that the suppression of \CIV\ absorption occurs relatively uniformly across \mbox{13.2 $\leq \log N <$ 15.0}. In the following sections we use these new measurements to investigate what fraction of the observed \OmCIV\ evolution could plausibly be driven by steady enrichment of the CGM/IGM by outflows (Section \ref{subsec:delayed_enrichment}) and by changes in the average ionization state of the absorbing gas (Section \ref{subsec:ionization_state}).

\subsection{Chemical Enrichment by Outflows}\label{subsec:delayed_enrichment}
\subsubsection{Toy Model}\label{subsubsec:toy_model}
We first explore the possibility that the rapid evolution in \OmCIV\ could trace early enrichment by galactic winds. Outflows transport metals liberated by supernovae and stellar winds from galaxies to the CGM/IGM, steadily increasing the metal content of galaxy halos over cosmic time. Our sample probes \CIV\ absorption at a time when the Universe is $\lesssim$~1.5 Gyr old; too young for a significant fraction of ejected metals to have been re-accreted onto galaxies (which occurs over a timescale of $\sim$~1~--~3 Gyr; \citealt{Oppenheimer09, Brook14, Uebler14}). Furthermore, the majority of galaxies ejecting metals at these redshifts have very shallow potential wells, increasing the likelihood that the metals will escape completely \citep[e.g.][]{Ferrara08, Pallottini14}. Therefore, it is reasonable to assume that the majority of metals that have been ejected remain outside of galaxies.

Following \citet[][hereafter \citetalias{Madau14}]{Madau14}, we assume that the comoving mass density of metals produced in, and ejected from, stars up to a given redshift $z$ is approximately proportional integral of the cosmic star formation rate density $\psi(z')$:
\begin{equation}\label{eqn:metal_density}
 \rho_{Z, \, \rm ejected}(z) \simeq y (1-R) \int_{0}^{t_{H(z)}} \psi(z') \left|\frac{dz'}{dt}\right| dt
\end{equation}
Here $1-R$ is the fraction of the initial stellar mass that is retained in stars, $y$ is the average mass of heavy elements created and ejected into the CGM/IGM per unit mass in those stars, and $t_{H(z)}$ is the age of the Universe at redshift $z$. The mass density of ejected carbon can be computed by multiplying Equation \ref{eqn:metal_density} by the mass fraction of metals in carbon, $A_C$:
\begin{equation}\label{eqn:carbon_density}
 \rho_{\rm C, \, ejected}(z) \simeq A_C \, \rho_{Z, \, \rm ejected}(z)
\end{equation}
Finally, the expected mass density of \CIV\ in the ejected material can be obtained by multiplying Equation \ref{eqn:carbon_density} by the fraction of carbon which is triply ionized:
\begin{equation}
 \rho_{\rm C IV}(z) \simeq \frac{\rm{C} \textsc{iv}}{{\rm C}_{\rm total}} \, \rho_{C, \, \rm ejected}(z)
\end{equation}
In this section we aim to quantify the expected evolution in \OmCIV\ originating from changes in the carbon content alone, so we assume that \CIV/C$_{\rm total}$ is independent of redshift. 

In order to compute the predicted \textit{absolute} evolution of \OmCIV\ over cosmic time, it is necessary to substitute in values for $y$, $R$, $A_C$, and \CIV/C$_{\rm total}$. However, the stellar population properties and \CIV\ ionization fraction are poorly constrained at these redshifts. We circumvent the uncertainties in these parameters by investigating whether the predicted \textit{rate of change} in the mean metal content is sufficient to explain the observed rate of evolution of \OmCIV, assuming that $y$, $R$ and $A_C$ do not vary significantly over the $\sim$~1.25 Gyr between the beginning of the Universe at $z\sim$~4.7. Specifically, we compare the predicted and observed fractional growth of \OmCIV\ between two redshifts $z_1$ and $z_2$:
\begin{equation}\label{eqn:xciv}
x_{\rm C IV}(z_1,z_2) = \frac{\rho_{\rm C IV}(z_1)}{\rho_{\rm C IV}(z_2)} \simeq \frac{\int_{0}^{t_{H(z_1)}} \psi(z') \left|\frac{dz'}{dt}\right| dt}{\int_{0}^{t_{H(z_2)}} \psi(z') \left|\frac{dz'}{dt}\right| dt}
\end{equation}

We emphasize that the calculation of $x_{\rm CIV}$ does not require us to adopt a specific value for the carbon yield or the fraction of produced carbon that is ejected into the CGM/IGM. However, it does require the extrapolation of $\psi(z')$ to \mbox{$z > 8$} where observational constraints are currently relatively limited. In the future, large surveys with JWST will enable much more robust measurements of $\psi(z')$ at these redshifts. We account for the uncertainty on $\psi(z')$ by performing our calculations using two parametrizations that differ significantly at $z\gtrsim$~8. The \citetalias{Madau14} parametrization has previously been used to calculate the cosmic mean metal content as a function of redshift \citep[e.g.][]{Peroux20}. However, constraints from the growing number of very high redshift galaxies observed with HST and JWST suggest that it over-predicts $\psi$ at $z\gtrsim$~8 \citep[e.g.][]{Mason15, Oesch18, Donnan22, Harikane22a, Harikane22b}. Early JWST results additionally suggest that distant galaxies are relatively blue, making it unlikely that a significant fraction of star-formation is obscured \citep[e.g.][]{Castellano22, Cullen22, Ferrara22, Finkelstein22, Santini22, Nanayakkara22}. We therefore also consider a second $\psi(z')$ parametrization published by \citet[][hereafter \citetalias{Harikane22a}]{Harikane22a} (their Equation 60) which assumes a constant star-formation efficiency at $z >$~10. This parametrization captures the observed rapid decline at $z\gtrsim$~6 but the scaling is too steep and \textit{under-estimates} the measured $\psi$ at $z\gtrsim$~10 \citep[e.g.][]{Bouwens22a, Bouwens22b, Harikane22b}. As a result, the $x_{\rm CIV}$ values implied by the \citetalias{Madau14} and \citetalias{Harikane22a} parametrizations may bracket the true fractional growth of \OmCIV\ as a result of chemical enrichment. 

Based on the \citetalias{Madau14} (\citetalias{Harikane22a}) parametrization for $\psi(z')$, the expected fractional growth in the mean carbon mass density between $z$ = 5.77 and $z$ = 4.72 (given by $x_{\rm C IV}(5.77, 4.72)$) is 2.1 (2.8). The \citetalias{Harikane22a} parametrization predicts a steeper rise in $\psi(z')$ at high redshift than the \citetalias{Madau14} model and therefore also predicts a faster growth in the cosmic mean metal content. However, in both cases the expected change in $\rho_{\rm CIV}$ is significantly smaller than the observed factor of 4.8 evolution in \OmCIV.

\begin{figure*}
 \includegraphics[scale=0.55, clip = True, trim = 0 0 5 0]{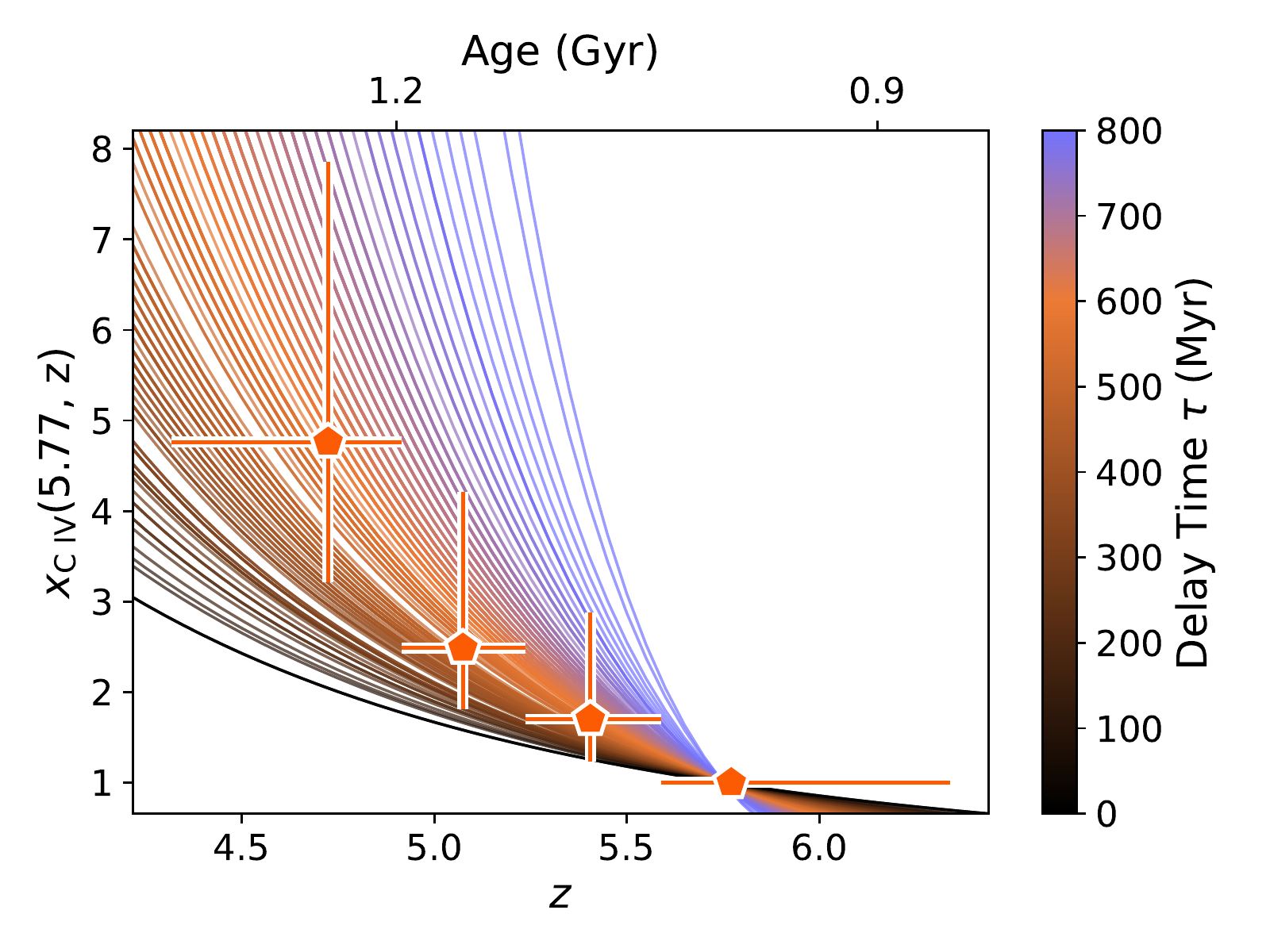} \includegraphics[scale=0.55, clip = True, trim = 0 0 5 0]{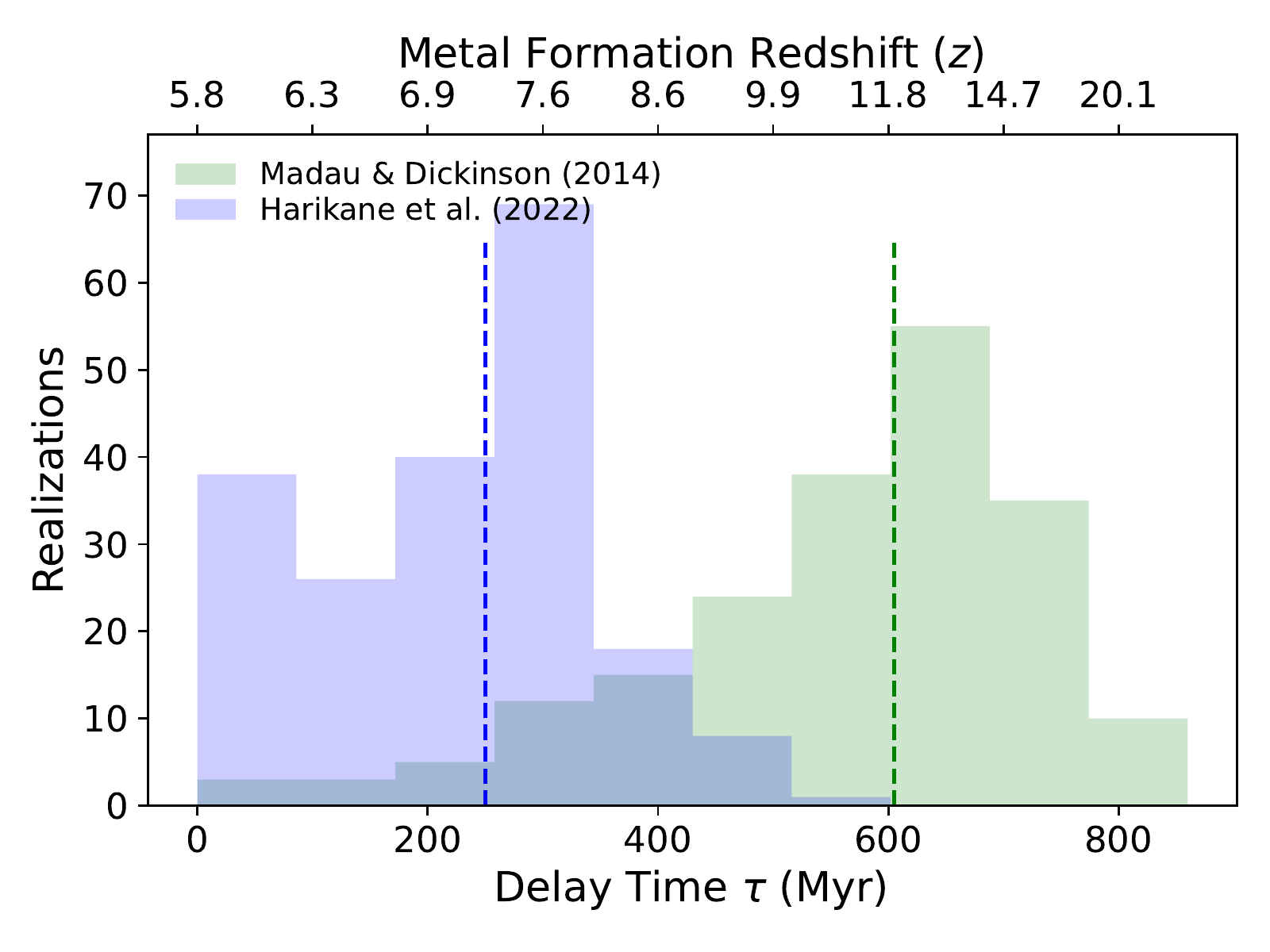}
\caption{Left: Measured $x_{\rm C IV}$ (see Equation \ref{eqn:xciv}) for absorbers with \mbox{13.2~$\leq \log N <$ 15.0} (orange pentagons), and the best-fit delayed enrichment curves for 200 bootstrap realizations of the measurements for the \citetalias{Madau14} parametrization of the cosmic star formation rate density $\psi(z')$ (colored curves). The color of each curve indicates the delay time of the corresponding best-fit model. The delayed enrichment model assumes that \OmCIV\ reflects the metal content of the CGM/IGM and scales as the integral of $\psi(z')$ with some delay time $\tau$ (see Equation \ref{eqn:delayed_enrichment}). Right: Distribution of best-fit delay time values for the 200 realizations, for both the \citetalias{Madau14} (green) and \citetalias{Harikane22a} (blue) parametrizations of $\psi(z')$. The dashed lines indicate the median values of the distributions (250~Myr and 605~Myr). The \citetalias{Madau14} parametrization has been shown to over-predict the star formation rate density at $z\gtrsim$~8 whilst the \citetalias{Harikane22a} predicts too little star formation, suggesting that the delay time required to explain the \OmCIV\ evolution by chemical enrichment alone likely lies between these two values.} \label{fig:delayed_enrichment} 
\end{figure*} 

This discrepancy may be resolved by accounting for the time delay between the formation of metals in stars and the deposition of these metals into the CGM/IGM where they are observed as \CIV\ absorbers. Low mass ($\lesssim$~8~M$_\odot$) stars eject significant quantities of metals during the asymptotic giant branch (AGB) phase, with a typical time delay of hundreds of Myr \citep[e.g.][]{Oppenheimer08, Kramer11}. It may then take a similar amount of time for outflows to transport these metals from the interstellar medium (ISM) to the sites of \CIV\ absorption which are typically located on scales of tens to hundreds of kpc \citep[e.g.][]{Garcia17a, Diaz21, Kashino22}. To account for this, we introduce a delay time $\tau$ into Equation \ref{eqn:xciv} such that the metal mass density of the \CIV-absorbing gas at a redshift $z$ is assumed to be proportional to the integral of $\psi(z')$ evaluated at \mbox{$t_{H(z)} - \tau$}:
\begin{equation}\label{eqn:delayed_enrichment}
x_{\rm C IV}(z_1,z_2,\tau) = \frac{\int_{0}^{t_{H(z_1)}-\tau} \psi(z') \left|\frac{dz'}{dt}\right| dt}{\int_{0}^{t_{H(z_2)}-\tau} \psi(z') \left|\frac{dz'}{dt}\right| dt}
\end{equation}

\subsubsection{Fitting and Physical Interpretation}
We investigate which delay time value best explains the observed \OmCIV\ evolution for each $\psi(z')$ parametrization as follows. We consider 90 different delay times spanning \mbox{$\tau$ = 0~--~900 Myr} with \mbox{$\Delta \tau$~=~10 Myr}. For each delay time $\tau$, we compute $x_{\rm C IV}$(5.77, $z$) for the four values of $z$ corresponding to the centres of our four redshift bins (given in Table \ref{table:measurements}), and then calculate the total $\chi^2$ difference between the four measured $x_{\rm C IV}$ values and the predicted values. The delay time yielding the smallest $\chi^2$ value is then selected as the best-fit value. The uncertainty on the best-fit delay time is estimated by repeating this calculation for all 200 bootstrap estimates of \OmCIV\ (see Section \ref{subsec:om_civ}). 

The $x_{\rm C IV}$ measurements are shown as orange pentagons in the left-hand panel of Figure \ref{fig:delayed_enrichment}. To illustrate the range of delay time values allowed by the bootstrap measurements we over-plot the best-fit delayed enrichment curves for the \citetalias{Madau14} parametrization, where the color represents the delay time of the best-fit model according to the colorbar to the right of the plot. In the right-hand panel we show the distribution of best-fit delay times for the 200 bootstrap iterations, for both the \citetalias{Madau14} (green) and \citetalias{Harikane22a} (blue) models. The upper $x$-axis shows the implied formation redshifts of the metals observed at $z\sim$~5.8. These plots are based on the \OmCIV\ values measured over \mbox{$\log N$ = 13.2 -- 15.0}, but consistent results are obtained for \OmCIV\ measured over \mbox{$\log N$ = 13.8 -- 15.0} (with larger errors due to the smaller sample size). 

We find that the observed \OmCIV\ evolution can be fully explained by gradual enrichment with a median best-fit delay time of 605~Myr for the \citetalias{Madau14} parametrization, or 250~Myr for the \citetalias{Harikane22a} parametrization. As discussed in Section \ref{subsubsec:toy_model}, these two parametrizations appear to bracket the true cosmic SFR density at $z\gtrsim$~10, and therefore it seems likely that the true delay time required to explain the \OmCIV\ evolution by chemical enrichment alone lies somewhere between these two values; i.e. \mbox{250 Myr $\lesssim \tau \lesssim$ 605 Myr}.

In this scenario, the metals observed in \CIV\ absorbers at $z\sim$~5.8 would have been formed at redshifts of \mbox{$z\sim$~7~--~12}; between $\sim$~100 Myr and $\sim$~600 Myr after the expected first light of galaxies at \mbox{$z\simeq$~15~--~20} \citep[e.g.][]{Hashimoto18, Labbe22, Naidu22, Yan22}. The abundance patterns of the highest redshift metal absorber at \mbox{$z$~=~6.84} suggest that early populations of \mbox{Population \textsc{ii}-like} stars were already present at $z\gtrsim$~10 \citep[e.g.][]{Simcoe20}. However, it is unclear whether it is physically reasonable to invoke such long delay times. 

We consider two sources of delay: the time taken for stars to return their carbon to the ISM, $\tau_{\rm ret}$, and the time taken for outflows to transport metals from the ISM to the CGM/IGM where they are observed as \CIV, $\tau_{\rm out}$. \citet{Kramer11} investigated the distribution of delay times between the formation of stars and the return of carbon to the ISM for a single age stellar population, including contributions from both Type II supernovae and AGB stars. They found that the mean delay time for a \citet{Chabrier03} IMF is \mbox{$\tau_{ret}$ = 160 Myr}\footnote{250~Myr for a \citet{Kroupa01} IMF}. Assuming that the outflow delay time $\tau_{\rm out}$ is given by $\tau_{\rm out} \simeq \tau - \tau_{\rm ret}$, our results imply that \mbox{$\tau_{\rm out} \simeq$~90~--~445~Myr}. 

The lower end of this $\tau_{\rm out}$ range is consistent with the predictions of \citet{Oppenheimer09} who find that simulated \CIV\ absorbers at $z\sim$~6 are typically observed \mbox{30~--~300~Myr} after being launched in a wind. $\tau_{\rm out}$ can also be expressed as the quotient of the characteristic distance from galaxies at which \CIV\ absorbers are detected (the impact parameter $D$) and the outflow velocity $v_{\rm out}$; i.e. \mbox{$\tau_{\rm out} = D / v_{\rm out}$}. This relationship can be used to infer the characteristic impact parameter of \CIV\ absorbers for a given delay time. Observations of \CII\ emission and rest-frame UV absorption suggest that star-forming galaxies at \mbox{$z\sim$~5~--~6} have typical outflow velocities of \mbox{300~--~500~\kms} \citep[e.g.][]{Gallerani18, Sugahara19, Ginolfi20a, Pizzati20}. Over travel times of \mbox{90~--~445~Myr}, such outflows would transport metals to typical distances of \mbox{$\lesssim$~30~--~230~kpc}. These values can be considered as upper limits because outflows may decelerate as they propagate \citep[e.g.][]{Nelson19}. \citet{Oppenheimer09} find that the typical velocities of winds associated with $z\sim$~6 \CIV\ absorbers, averaged over their lifetimes, are $\sim$~50\% of their launch velocities. Accounting for this deceleration suggests that the ejected metals would more likely reach impact parameters of \mbox{$\sim$~15~--~115~kpc}. 

The lower half of this impact parameter range is consistent with predictions from simulations and with observed values. \citet{Oppenheimer09} predict that most \CIV\ absorbers at $z\sim$~6 trace metals on their first journey into the IGM, making it simpler to link absorbers with galaxies at this time compared to later epochs. They find that most simulated \CIV\ absorbers at this redshift lie $\sim$~5~--~50 kpc from their host galaxies. Observationally linking galaxies to absorbers at these redshifts is very challenging because dwarf galaxies are expected to play a significant role in the enrichment of the CGM/IGM \citep[e.g.][]{Garcia17a, Diaz21}. However, several studies have identified galaxies within \mbox{$\sim$~50 kpc} of strong \CIV\ absorbers. \citet{Cai17} find a \Lya\ emitter (LAE) 42 kpc from a \CIV\ absorber with \mbox{$\log N >$ 14.0} at $z\sim$~5.74. \citet{Diaz21} find that 2/8 of the faint LAEs associated with CIV absorbers with \mbox{$\log N >$ 13.5} at \mbox{4.9 $\leq z \leq$ 5.7} have impact parameters of $<$~50~kpc. \citet{Codoreanu18} find that \CIV\ systems with \mbox{$\log N >$ 14} are preferentially associated with low-ionization counterparts and may therefore trace multi-phase gas at the interface between the CGM and the IGM. Some strong \CIV\ absorbers do not appear to have any galaxies in this impact parameter range but instead show one or more galaxies at larger distances, suggesting that the absorbing gas could be part of an extended structure connecting galaxies \citep[e.g.][]{Diaz21}. However, the presence of intrinsically faint or dust-obscured galaxies below the detection limit at smaller impact parameters cannot be ruled out \citep[e.g.][]{Neeleman20}. Deeper searches for galaxies associated with \CIV\ absorbers are required to obtain better constraints on their typical impact parameters.

Together, these findings suggest that the observed rapid evolution in \OmCIV\ over \mbox{4.3 $\lesssim z \lesssim$ 6.3} could plausibly be explained by gradual chemical enrichment with a delay time of \mbox{$\sim$~250~--~605 Myr}, comprising $\sim$~160 Myr for the release of carbon from AGB stars and \mbox{$\sim$~90~--~445 Myr} for the transport of carbon in outflows. Assuming lifetime-averaged outflow velocities of \mbox{$\sim$~150~--~250~\kms}, the expected typical impact parameter for \CIV\ absorbers would be \mbox{$\sim$~15~--~115~kpc}, which is consistent with observed and predicted values.

The orange dashed lines in Figure \ref{fig:omciv} show the predicted \OmCIV\ evolution across \mbox{4 $\lesssim z \lesssim$ 7} for the best-fit \citetalias{Madau14} model. (By design, the curve representing the best-fit \citetalias{Harikane22a} model is almost the same). If the carbon enrichment of the CGM progresses in this manner, then \CIV\ absorption should become increasingly rare at $z >$~6, which is consistent with the scarcity of detections at these redshifts. \citet{Cooper19} and \citet{Simcoe20} find no evidence of \CIV\ in the earliest known intervening metal absorber at \mbox{$z$~=~6.84}, and \citet{Bosman17} measured an upper limit on \OmCIV\ at $z\sim$~6.5 from a single \mbox{$z\sim$~7} quasar sightline (see left panel of Figure \ref{fig:omciv}).

However, we note that this model cannot explain the observed evolution in \OmCIV\ at $z < 4$. The model predicts that \OmCIV\ should continue to increase rapidly at \mbox{$z \lesssim$~5}; in tension with the measurements which show that \OmCIV\ changes very little between $z\sim$~4.7 and $z\sim$~3.4. As mentioned in Section \ref{subsubsec:toy_model}, the toy model assumes that the current metal content of the CGM/IGM represents the accumulation of all metals ejected from galaxies in the past, and this assumption starts to break down when the Universe becomes old enough that a non-negligible fraction of the ejected metals have been re-accreted onto galaxies. However, it seems unlikely that metal recycling would lead to such a pronounced change in the growth rate of \OmCIV\ between $z\sim$~5 and $z\sim$~4. In Section \ref{subsubsec:ionization_state_obs} we discuss the possibility that this transition may be more naturally explained by changes in the ionization state of the absorbing gas.

\subsubsection{Caveats}
The calculations presented here are based on a toy model with many associated uncertainties, some of which have already been discussed. The $\psi(z)$ measurements on which the \citetalias{Madau14} and \citetalias{Harikane22a} parametrizations are based only include contributions from galaxies with \mbox{$M_{UV} < -17$}, corresponding to \mbox{SFR $\gtrsim$ 0.3 M$_\odot$ yr$^{-1}$}. However, fainter galaxies have shallower gravitational potentials and may therefore contribute substantially to the enrichment of the CGM/IGM \citep[e.g.][]{Garcia17a, Diaz21}.

We assume that the carbon yields of stars remain constant between the beginning of the Universe and $z\sim$~4.7. However, studies of metal-poor stars in the Milky Way have revealed a significant population of Carbon Enhanced Metal Poor (CEMP) stars (see \citealt{Beers05} for a review), and there is evidence for enhanced [C/O] ratios in the most metal-poor stars and metal absorbers \citep[e.g.][]{Cooke17, Banados19}. If stellar populations at $z\sim$~6 have enhanced carbon yields compared to stellar populations at $z\sim$~5, the carbon content would grow more slowly than the integral of $\psi(z')$, requiring larger delay times to explain the observed \OmCIV\ evolution which could become unrealistic given the age of the Universe at these redshifts ($\sim$~1~Gyr at $z\sim$~6). 

\subsection{Ionization State of the CGM}\label{subsec:ionization_state}
We also consider the possibility that the rapid evolution in \OmCIV\ could be driven by a change in the ionization state of the CGM/IGM. It is unclear whether gas in the halos of \mbox{$z\sim$~5~--~6} galaxies is primarily ionized by the UV background or by radiation from the host galaxy. As discussed earlier, simulations often under-predict the incidence of \CIV\ absorbers at $z\gtrsim$~5, which could be an indication that galaxies self-ionize their halos \citep[e.g.][]{Keating16}. If the halo gas is primarily ionized by the local radiation field, then the ionization state of the CGM/IGM would be expected to correlate with the typical strength and hardness of this radiation field.

However, observational evidence suggests that the UV background is the primary source of ionization in galaxy halos. \citet{Lau16} found that the ionization parameter of the gas around quasar host galaxies at \mbox{$z\sim$~2~--~3} increases towards larger radii, suggesting that the halo gas is primarily ionized by the UV background. \citet{Meyer19} found evidence for excess \Lya\ transmission in the vicinity of \CIV\ absorbers at \mbox{$z\sim$~5~--~6}, suggesting that these absorbers trace local enhancements in the UV background \citep[see also][]{Finlator16}. If the UV background is the dominant source of ionization, then the properties of metal absorbers may be significantly impacted by the reionization of hydrogen and/or \HeII.

\subsubsection{\HeII\ Reionization}
The energy required to ionize \HeII\ (54.4 eV) lies between the energies to ionize \CIII\ and \CIV\ (47.9 and 64.5 eV, respectively), and therefore \HeII\ reionization may impact the ionization states of carbon absorbers. \HeII\ reionization is expected to end at around $z\sim$~3 \citep[e.g.][]{DaviesF14, UptonSanderbeck16, Worseck19, Makan21}. \citet{Songaila98} reported a jump in the \SiIV/\CIV\ ratio at $z\sim$~3 that they attributed to a change in the UVB associated with the end of \HeII\ reionization, but subsequent studies did not reproduce this finding \citep[e.g.][]{Kim02, Boksenberg03, Cooksey11, Boksenberg15}. The onset of \HeII\ reionization is uncertain but may have occurred around $z\sim$~4.5, when the temperature of the IGM has been observed to rise \citep[e.g][]{Becker11, Boera14, Walther19, Gaikwad21}. This is intriguingly similar to the redshift where we observe a transition in the rate of evolution of \OmCIV. It could be that the properties of \CIV\ absorbers are impacted by changes in the relative contribution of QSOs to the UV background and the mean free path of \HeII\ ionizing photons during \HeII\ reionization. However, the incidence rate of \SiIV\ (which is not sensitive to the \HeII\ break in the UV background) follows the same trends as \CIV\ \citep{DOdorico22}. Furthermore, the \CIV\ incidence rate changes very little over $z\sim$~3~--~4.5 when the \HeII\ filling factor of the IGM must be changing rapidly. For these reasons, we suggest that \HeII\ reionization is likely not a key factor driving the observed evolution in \OmCIV.

\subsubsection{Hydrogen Reionization}
The end of hydrogen reionization is expected to be marked by rapid changes in the amplitude of the UV background \citep[e.g.][]{Kulkarni19, Keating20, Garaldi22, Lewis22}, driving a transition of metals from primarily neutral or low-ionization environments to more highly ionized environments once reionization is complete \citep[e.g.][]{Becker19}. There are multiple independent lines of evidence suggesting that signatures of reionization persist below $z\sim$~6 \citep[e.g.][]{Zhu21, Bosman22}, indicating that this effect could be significant over the redshift range probed in this study. Indeed, there is growing observational evidence for a shift in the typical ionization environments of metal absorbers at \mbox{$z\sim$~5~--~6}. This work confirms previous findings that the incidence of \CIV\ absorbers declines rapidly at $z\gtrsim$~4.9, and similar behaviour is observed for \SiIV\ \citep[e.g.][]{DOdorico22}. Examining the evolution of low-ionization ions can help to deduce whether the decline in high-ionization absorbers is primarily driven by ionization or metal content. The incidence of weak \MgII\ absorbers remains approximately constant over the probed redshift range \citep[e.g.][]{Bosman17, Chen17, Codoreanu17, Zou21}, and the incidence of weak \OI\ absorbers \textit{increases} over \mbox{4.1 $< z <$ 6.5} \citep[e.g.][]{Becker19}. The cosmic mean metallicity decreases towards higher redshift, so this upturn can only be explained by an increase in the fraction of oxygen observed as \OI\ \citep[e.g.][]{Doughty19} and/or an increase in the relative oxygen abundance (see e.g. \citealt{Welsh22} and references therein). 

It is interesting to consider whether both the upturn in weak (\mbox{$W < 0.2$\AA} rest-frame) \OI\ absorbers at $z\gtrsim$~5.7 and the decline in \CIV\ absorption at $z\gtrsim$~5 can be explained by a single population of absorbers transitioning from higher ionization environments at $z\sim$~5 to lower ionization environments at $z\sim$~6. In this scenario, if weak \CII\ absorbers follow the same trend as weak \OI\ absorbers, then we would expect the decline in \CIV\ absorption to be dominated by systems with $W < 0.2$\AA, which corresponds to an apparent column density of \mbox{$\log N_a$(\CIV) $\lesssim$~13.7} (assuming the absorption lies on the linear part of the curve of growth; \citealt{Spitzer78}). However, we have shown that the slope of the \CIV\ CDDF does not vary as a function of redshift (see Section \ref{subsec:cddf} and Figure \ref{fig:cddf}), indicating that the strong and weak \CIV\ absorbers decrease in incidence at similar rates. We caution that the measured number densities of strong ($W > 0.2$\AA) \OI\ absorbers are relatively uncertain due to small number statistics and do not preclude the possibility of an upturn at $z\gtrsim$~5.7 \citep[e.g.][]{Becker19}. The interpretation is additionally complicated by the fact that \OI\ probes dense gas whereas \CIV\ probes diffuse gas which in lower ionization environments may manifest as \CIII~$\lambda$977\AA\ (e.g. \citealt{Cooper19}) which falls in the saturated \Lya\ forest.

\subsubsection{Constraints from Measurements of \OmCII}\label{subsubsec:ionization_state_obs}
Changes in the ionization state of the absorbing gas can be examined more directly by measuring the column density ratios of different ions of the same species (e.g. \mbox{$\log N$(\CII)/$\log N$(\CIV)}) which are insensitive to changes in metallicity and/or abundance patterns over cosmic time. Previous studies have found evidence for an increase in the \CII/\CIV\ ratio at $z\gtrsim$~5.7 \citep[e.g.][]{DOdorico13, Cooper19, Simcoe20}, suggesting that some portion of the observed \OmCIV\ evolution could be the result of carbon transitioning to lower ionization states. 

We use the E-XQR-30 metal absorber catalog to compute $\Omega_{\rm C \textsc{II}}$ in the same redshift bins as our \OmCIV\ measurements and investigate whether the decrease in \OmCIV\ is accompanied by an increase in the frequency of carbon in lower ionization states. We select intervening \CII\ absorbers from the primary absorber sample using the method described in Section \ref{subsec:sample} and consider systems in the same $\log N$ range as the \CIV\ absorbers\footnote{We note that the sample completeness exceeds 50\% for \CII\ absorbers with \mbox{13.2 $\leq \log N <$ 15}; see \citetalias{Davies22Survey}.}. We do not present measurements of \OmCII\ for the two lowest redshift bins where \CII\ is inaccessible due to the saturation of the \Lya\ forest. In a forthcoming paper we will investigate the use of \MgII\ as a proxy for \CII\ \citep[e.g.][]{Cooper19} to place similar constraints on the ionization state of the absorbing gas in the lower redshift bins (Sebastian et al., in prep).

\begin{figure}
 \includegraphics[scale=0.55]{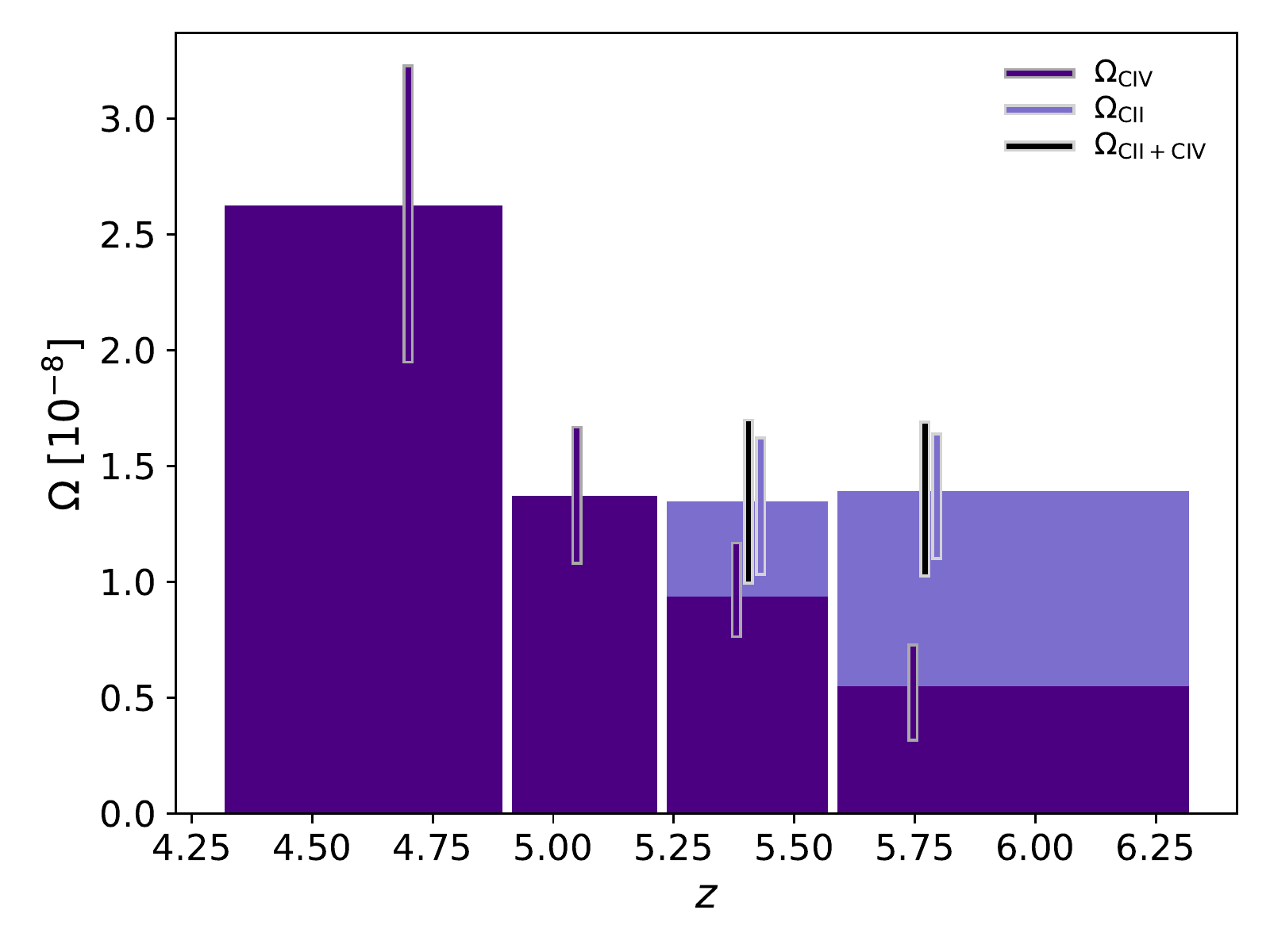} 
\caption{Bar graph illustrating the contribution of \CIV\ and \CII\ to the cosmic mass density of carbon in each redshift bin. Vertical lines illustrate the measurement errors and are centered on the path-length-weighted mean redshift of each bin with small offsets added for clarity. We are unable to measure \OmCII\ in the two lowest redshift bins due to the saturation of the \Lya\ forest. The evolving balance between \OmCII\ and \OmCIV\ across the two highest redshift bins suggests that changes in the UV background driven by hydrogen reionization may contribute to the decline in \OmCIV.} \label{fig:carbon_ion_fractions} 
\end{figure}

Figure \ref{fig:carbon_ion_fractions} illustrates the contributions of \CIV\ and \CII\ (when available) to the cosmic mass density of carbon over the four redshift bins used in this paper. Over the two highest redshift bins, the decrease in \OmCIV\ is completely balanced by an increase in \OmCII. At $z\sim$~5.8 we measure \mbox{$\Omega_{\rm C \textsc{II}}$~=~(8.41$^{+2.29}_{-2.73}$)~$\times$~10$^{-9}$} and \mbox{$\Omega_{\rm C \textsc{II} + C \textsc{IV}}$~=~(1.39$^{+0.29}_{-0.38}$)~$\times$~10$^{-8}$}, and at $z\sim$~5.4 we measure \mbox{$\Omega_{\rm C \textsc{II}}$~=~(4.11$^{+2.56}_{-2.97}$)~$\times$~10$^{-9}$} and \mbox{$\Omega_{\rm C \textsc{II} + C \textsc{IV}}$~=~(1.35$^{+0.30}_{-0.35}$)~$\times$~10$^{-8}$}. This suggests that changes in the typical ionization state of carbon as a result of reionization may play a significant role in driving the evolution of \OmCIV\ at the highest redshifts probed by our sample. 

On the other hand, changes in \OmCII\ cannot balance the evolution in \OmCIV\ at lower redshifts. The combined $\Omega_{\rm C \textsc{II} + C \textsc{IV}}$ measured at $z\sim$~5.4 is smaller than the \OmCIV\ measured at $z\sim$~4.7, indicating that $\Omega_{\rm C \textsc{II} + C \textsc{IV}}$ must increase over this redshift range. We note that 91\% (81\%) of the observed evolution in \dndx\ (\OmCIV) occurs over this interval. These results suggest that a change in the typical ionization state of metal absorbers is likely not the only factor driving the evolution in \OmCIV\ over \mbox{4.3 $\lesssim z \lesssim$ 6.3}.

The main limitation of this analysis is that we are only able to observe two ionization states of carbon. It is unclear what fraction of the missing \CIV\ might be present in \CIII\ which is often the dominant ionization state at $z <$~1 \citep[e.g.][]{Lehner18} but cannot be observed for any of the objects in our study because \CIII~$\lambda$977\AA\ falls in the saturated \Lya\ forest. Many simulations predict that the overall fraction of carbon in \CIV\ changes relatively little over the probed redshift interval \citep[e.g.][]{Oppenheimer09, Cen11, Finlator15, Rahmati16, Garcia17b}. However, some of these simulations were run using a uniform UV background which has been shown to be inconsistent with observations of the \Lya\ forest at $z\geq$~5.3 \citep{Bosman22}. The simulations that do model an inhomogenous reionization are computationally limited to volumes smaller than the spatial scales over which the \Lya\ forest opacities have been observed to fluctuate \citep{Becker15} due to the need to simultaneously model galaxy evolution processes on orders-of-magnitude smaller scales. It therefore remains unclear whether or not cosmological simulations predict a strong evolution in the fraction of carbon observed as \CIV\ across this redshift range.

\subsubsection{Physical Intepretation}
Changes in the ionization state of the absorbing gas may offer a natural explanation for the observed two-stage evolution of \OmCIV, characterized by a rapid rise in \OmCIV\ over \mbox{6 $\gtrsim z \gtrsim$ 5} followed by a much more gradual increase over \mbox{5 $\gtrsim z \gtrsim$ 1.5}. \citet{Bosman22} found that spatial fluctuations in the UV background and/or the properties of the IGM are required to explain the scatter in the \Lya\ optical depth down to \mbox{$z$~=~5.3}. If these fluctuations are associated with the end of hydrogen reionization, then we would expect to observe a large change in the amplitude of the UV background at similar redshifts \citep[e.g.][]{Kulkarni19, Keating20, Garaldi22, Lewis22}, which in turn could contribute to the observed rapid evolution of \OmCIV\ in the early Universe. The flatter evolution in \OmCIV\ observed at $z \lesssim$~5 would then reflect the more gradual changes in the UV background following the end of reionization.

\begin{figure}
 \includegraphics[scale=0.55]{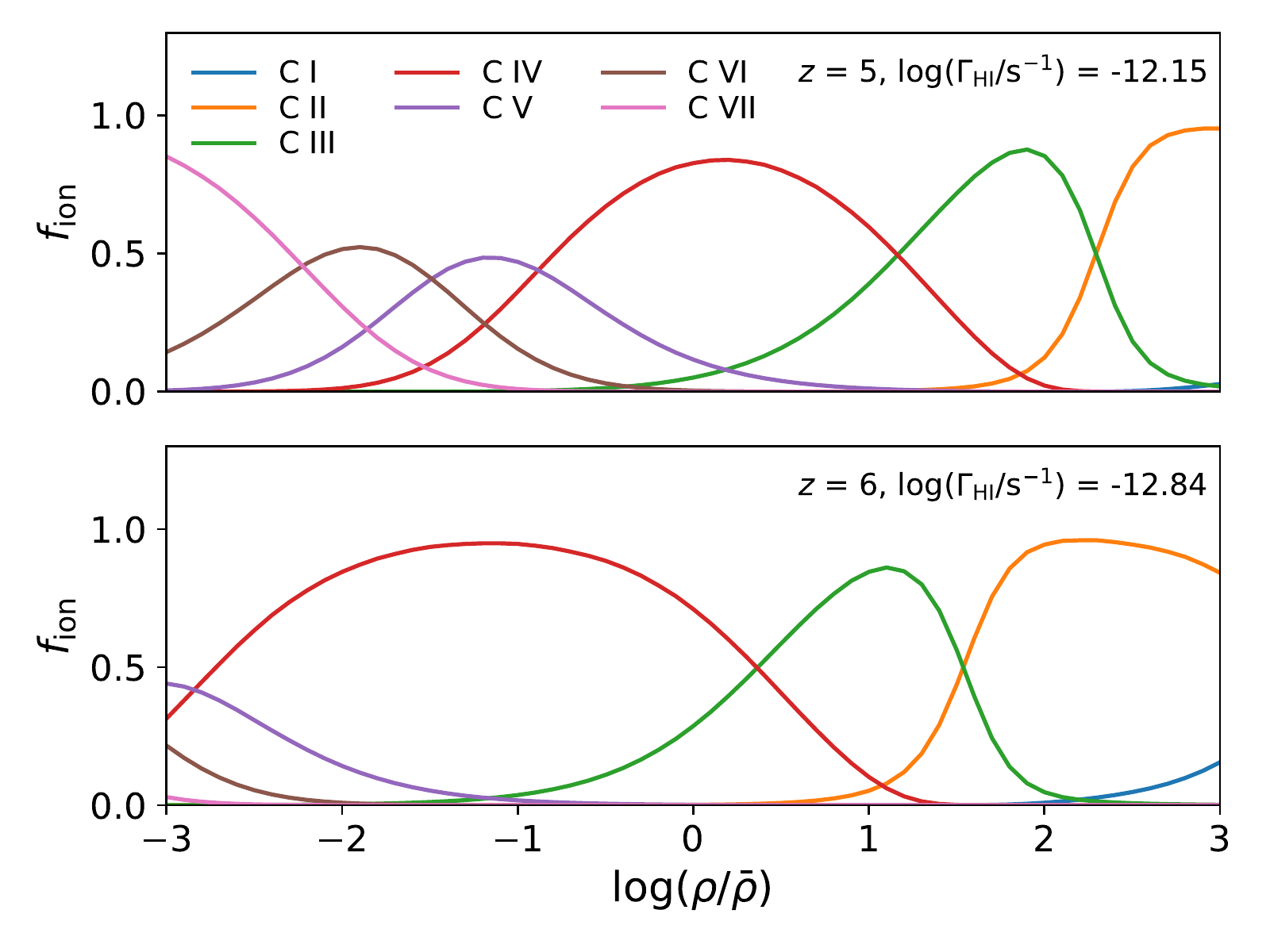} 
\caption{Illustration of the possible ranges of gas densities probed by different ionization states of carbon at $z$ = 5 and $z$ = 6. The curves are the outputs of \textsc{cloudy} photoionization models. We fix the gas temperature to 10$^4$~K and adopt the 2011 update of the \citet{FaucherGiguere09} ionizing spectrum with a density-dependent correction for gas self-shielding as described in \citet{Keating16}. The ionizing spectrum is re-scaled at each redshift to match the H~\textsc{i} photoionization rates measured by \citet{Calverley11}.} \label{fig:cloudy_modelling} 
\end{figure}

To illustrate how a reduction in the amplitude of the UV background could lead to a decline in \OmCIV, we performed simple photoionization modelling with \textsc{cloudy} \citep{Ferland17} to estimate the gas densities probed by different carbon ions at $z\sim$~5 and $z\sim$~6. The gas temperature is fixed at \mbox{$T$ = 10$^4$~K} which is well below the collisional ionization temperature for \CIV. We adopt the 2011 update of the \citet{FaucherGiguere09} ionizing spectrum, with a density-dependent correction for gas self-shielding as described in \citet{Keating16}. The amplitude of the ionizing spectrum is re-scaled to match the H~\textsc{i} photoionization rate measurements from \citet{Calverley11}\footnote{\mbox{$\Gamma_{\rm HI}$ = $10^{-12.15}$ s$^{-1}$} at \mbox{$z$ = 5} and \mbox{$\Gamma_{\rm HI}$ = 10$^{-12.84}$ s$^{-1}$ at $z$ = 6}. This evolution is quite steep but is reproduced in some simulations of a late-ending reionization \citep[e.g.][]{Lewis22}.}. The shape and amplitude of the UV background are difficult to measure at these redshifts and may also vary spatially \citep[e.g.][]{DaviesF16}. We emphasize that the adopted values are used only to illustrate how changes in the strength of the UV background translate to changes in the distribution of carbon between ionization states.

The results are shown in Figure \ref{fig:cloudy_modelling}. As the amplitude of the UV background decreases (moving from $z\sim$~5 to $z\sim$~6), \CIV\ absorption (shown in red) becomes increasingly sensitive to diffuse gas which may be located further from the galaxies responsible for driving outflows and may therefore be less metal-enriched, helping to explain the observed decline in \OmCIV. The highest density material probed by \CIV\ absorption at $z\sim$~5 is predominately traced by \CIII\ (green) at $z\sim$~6, highlighting the importance of considering the unobservable \CIII\ absorption when investigating the redshift evolution of \OmCIV. It is also clear that the range of densities probed by \CII\ absorption (orange) extends towards lower values at $z\sim$~6 than at $z\sim$~5. These effects may partially explain the observed decline in the \OmCIV/\OmCII\ ratio over the two highest redshift bins of our sample (Figure \ref{fig:carbon_ion_fractions}). We note that the shape of the ionizing spectrum may evolve simultaneously with the amplitude between $z\sim$~6 and $z\sim$~5 \citep[e.g.][]{Finlator18}. This effect is neglected here, but would likely enhance the trends seen in Figure \ref{fig:cloudy_modelling}.

It is not straightforward to determine how changes in the UV background would impact the slope of the \CIV\ CDDF. \citet{Keating16} found that the relative number of strong and weak \CIV\ absorbers is sensitive to both the shape and amplitude of the UV background. However, they did not compute their simulated CDDF and quantify the changes in its slope. \citet{Finlator16} found that any changes in the CDDF slope for different UV background models would be too small to detect given current measurement uncertainties, even for an extreme model where the UV background is completely dominated by QSOs. More theoretical investigation is therefore required to understand the impact of the evolving amplitude and shape of the UV background in the context of a late-ending reionization, and whether this would produce a detectable signature in the slope of the CDDF.

\section{Summary}\label{sec:conclusions}
We use a sample of 260 \CIV\ absorbers at \mbox{4.3 $\lesssim z \lesssim$ 6.3} to examine the rate and physical origin of the rapid decline in the cosmic mass density of \CIV\ (\OmCIV) at $z\gtrsim$~5. The absorbers were drawn from the catalog published by \citet{Davies22Survey} who performed a systematic search for metal absorbers in high S/N XSHOOTER spectra of 42 $z\sim$~6 quasars, of which 30 are drawn from the ESO large program \mbox{XQR-30}. The sample contains a factor of $\sim$~3 more \CIV\ absorbers at $z >$~5 than any previous compendium in the literature, enabling us to make the most robust measurements to date of \CIV\ absorber statistics in the high redshift Universe.

We measure the number density (\dndx) and cosmic mass density (\OmCIV) of \CIV\ absorption in four redshift bins. The sample is $\geq$~50\% complete for \mbox{$\log N \geq$ 13.2}, and our analysis is focused on absorbers in the column density range \mbox{13.2 $\leq \log N <$ 15.0}. Both \dndx\ and \OmCIV\ decline rapidly but smoothly over the probed redshift range, with \dndx\ decreasing by a factor of 3.4~$\pm$~0.9 and \OmCIV\ decreasing by a factor of 4.8~$\pm$~2.0 over this $\sim$~300~Myr period. This decline is much more rapid than what is observed over \mbox{1.5 $< z <$ 4.0} for \CIV\ absorbers from the \citet{DOdorico10} sample. Interestingly, the statistics of \CIV\ absorbers at $z\sim$~4.7 are comparable to those at $z\sim$~3.4. Our results are consistent with previous findings that the incidence of strong \CIV\ absorbers is approximately constant over \mbox{1.5 $\lesssim z \lesssim$ 5} and declines rapidly towards higher redshifts. However, the larger sample size used in this work enables us to measure \dndx\ and \OmCIV\ in smaller redshift intervals and therefore constrain the rate and timing of the decline to higher precision than previous studies.

We investigate whether the observed evolution is dominated by absorbers in a particular column density range by fitting the \CIV\ column density distribution function (CDDF) in two redshift bins centered at $z\sim$~5.6 and $z\sim$~4.9. The normalization decreases by a factor of 2.4~$\pm$~0.3, consistent with the observed number density evolution, but there is no evidence for a change in slope between the two redshift bins \mbox{($\Delta \alpha$ = $-$0.01 $\pm$ 0.17)}. Our work provides the first robust indication that the decline in \CIV\ absorption occurs approximately uniformly across \mbox{13.2 $\leq \log N <$ 15.0} and is not preferentially driven by strong or weak absorbers.

We use our improved measurements to examine whether the rapid evolution in \OmCIV\ could reflect a change in the carbon content of the CGM/IGM driven by the fast production of metals in the high redshift Universe. Assuming that the mean metal content of the CGM/IGM scales with the integral of the cosmic star formation rate density, it would be expected to increase by a factor of $<$~2.8 between $z\sim$~5.8 and $z\sim$~4.7, which is less than 60\% of the observed \OmCIV\ evolution. If we additionally account for the time taken to deposit carbon into the ISM and eject it into galaxy halos \mbox{($\sim$~250~--~605~Myr)}, we find that chemical evolution could plausibly explain all of the observed evolution in \OmCIV\ over \mbox{4.3 $\lesssim z \lesssim$ 6.3}. However, this model cannot easily explain why \OmCIV\ declines much more slowly over \mbox{1.5 $\lesssim z \lesssim$ 5} than at $z\gtrsim$~5.

Finally, we explore whether the \OmCIV\ evolution could be driven by a change in the fraction of carbon observed as \CIV\ by comparing the relative evolution of \OmCIV\ and \OmCII. Examining two ions of the same element removes uncertainties associated with variations in metallicity and/or abundance patterns over cosmic time. Over the two highest redshift bins ($z\sim$~5.4 and $z\sim$~5.8), we find that the decrease in \OmCIV\ is completely balanced by an increase in \OmCII, suggesting that the decline in \OmCIV\ over this redshift range may trace a decrease in the typical ionization state of carbon as the IGM becomes more neutral. In contrast, the combined \mbox{\OmCIV~+~\OmCII} measured at $z\sim$~5.4 is smaller than \OmCIV\ alone at $z\sim$~4.7. This could indicate that a change in the typical ionization state of metal absorbers is not the dominant factor driving the evolution in \OmCIV\ over \mbox{4.3 $\lesssim z \lesssim$ 6.3}. However, we are unable to measure the contribution from \CIII\ which is inaccessible at these redshifts. We note that rapid changes in the amplitude and/or hardness of the UV background as a result of hydrogen reionization may offer a natural explanation for the increased rate of decline in \OmCIV\ at $z\gtrsim$~5.

Much of this interpretation is based on toy model estimates for the gas-phase metallicity and measurements of the \OmCII/\OmCIV\ ratio as a proxy for the ionization state of the absorbing gas. In the future, photoionization modelling of column density measurements for ions covering a wide range of ionization states could be performed to obtain probability distributions for the metallicity and ionization parameter of the absorbing gas at $z\sim$~5 and $z\sim$~6 given different assumptions for the strength and shape of the ionizing photon background \citep[e.g.][]{Glidden16, Cooper19, Simcoe20}. Furthermore, carbon emission lines (including CO, [C~\textsc{ii}] and C~\textsc{iii}]) are a complementary probe of carbon in early galaxies. Combining emission line measurements with our absorption line results would provide further constraints on the evolution of the total carbon abundance in the early Universe. 

Overall, the new measurements presented in this paper suggest that gradual chemical enrichment by outflows could plausibly explain the observed rapid evolution in \OmCIV\ over \mbox{4.3 $\lesssim z \lesssim$ 6.3} without any need for changes in the ionization state of the absorbing gas. However, the increased incidence of \CII\ and \OI\ absorbers in the early Universe \citep[see also e.g.][]{Becker19, Cooper19} combined with the steepening of the decline in \OmCIV\ at $z\gtrsim$~5 suggest that reionization may also have a significant impact on the evolution of \CIV\ absorbers over cosmic time.

\section*{Acknowledgements}
We thank the referee for their valuable suggestions which improved the clarity of this paper. We thank Kristian Finlator for providing the statistics of \CIV\ absorbers in the \textsc{Technicolor Dawn} simulation and for constructive feedback on the draft manuscript. RLD acknowledges the support of a Gruber Foundation Fellowship research support grant. This research was supported by the Australian Research Council Centre of Excellence for All Sky Astrophysics in 3 Dimensions (ASTRO 3D), through project number CE170100013. SEIB and RAM acknowledge funding from the European Research Council (ERC) under the European Union's Horizon 2020 research and innovation programme (grant agreement no. 740246 "Cosmic Gas''). LCK was supported by the European Union’s Horizon 2020 research and innovation programme under the Marie Skłodowska-Curie grant agreement No. 885990. ACE acknowledges support by NASA through the NASA Hubble Fellowship grant $\#$HF2-51434 awarded by the Space Telescope Science Institute, which is operated by the Association of Universities for Research in Astronomy, Inc., for NASA, under contract NAS5-26555. EPF is supported by the international Gemini Observatory, a program of NSF’s NOIRLab, which is managed by the Association of Universities for Research in Astronomy (AURA) under a cooperative agreement with the National Science Foundation, on behalf of the Gemini partnership of Argentina, Brazil, Canada, Chile, the Republic of Korea, and the United States of America. AP acknowledges support from the ERC Advanced Grant INTERSTELLAR H2020/740120. Based on observations collected at the European Organisation for Astronomical Research in the Southern Hemisphere under ESO Programme IDs 0100.A-0625, 0101.B-0272, 0102.A-0154, 0102.A-0478, 084.A-0360(A), 084.A-0390(A), 084.A-0550(A), 085.A-0299(A), 086.A-0162(A), 086.A-0574(A),087.A-0607(A), 088.A-0897(A), 091.C-0934(B),  096.A-0095(A), 096.A-0418(A), 097.B-1070(A), 098.B-0537, 098.B-0537(A), 1103.A-0817, 294.A-5031(B), 60.A-9024(A). This research made use of NASA's Astrophysics Data System, as well as \textsc{Astrocook} \citep{Cupani20}, \textsc{Astropy} \citep{Astropy13, Astropy18}, \textsc{Matplotlib} \citep{Hunter07}, \textsc{Numpy} \citep{Harris20}, and \textsc{Scipy} \citep{Scipy20}. 

\section*{Data Availability}
The metal absorber catalog used in this paper and the \textsc{python} script used to calculate the absorption path lengths are publicly available and can be downloaded from this GitHub repository: \href{https://github.com/XQR-30/Metal-catalogue}{\url{https://github.com/XQR-30/Metal-catalogue}}.

\bibliography{../bibliography/mybib}

\appendix

\label{lastpage}
\end{document}